\documentstyle[prl,multicol,aps]{revtex}

\newcommand{\BEQ}{\begin{equation}}
\newcommand{\EEQ}{\end{equation}}
\newcommand{\BEA}{\begin{eqnarray}}
\newcommand{\EEA}{\end{eqnarray}}
\newcommand{\nn}{\nonumber \\}
\renewcommand{\d}{{\rm d}}

\newcommand{\th}{\theta }

\newcommand{\half}{\frac{1}{2}}
\newcommand{\la}{\lambda}

\newcommand{\BP}{P_{0}}
\newcommand{\BX}{X_{0}}
\newcommand{\Bx}{x_{0}}
\newcommand{\Bp}{p_{0}}
\newcommand{\R}{{\cal R}}

\newcommand{\RA}{{\rm A}}
\newcommand{\RS}{{\rm S}}
\newcommand{\re}{{\rm e}}

\newcommand{\N}{{\cal N}}
\newcommand{\V}{{\cal V}}
\newcommand{\W}{{\cal W}}

\newcommand{\ri}{{\rm i}}
\newcommand{\dt}{\partial _{t}}
\newcommand{\dP}{\partial _{P}}
\newcommand{\dX}{\partial _{X}}
\newcommand{\HX}{\hat{X}}
\newcommand{\HP}{\hat{P}}
\newcommand{\HS}{\hat{S}}
\renewcommand{\thesection}{\arabic{section}}

\renewcommand{\theequation}{\thesection\arabic{equation}}

\draft

\begin{document} 
\draft
\title{Quantum measurement as driven phase transition:
An exactly solvable model. }
\author{Armen E. Allahverdyan$^{1,3)}$, Roger Balian$^{1)}$ 
and Theo M. Nieuwenhuizen$^{2)}$}
\address{$^{1)}$ S.Ph.T., CEA Saclay, 91191 Gif-sur-Yvette cedex, France\\
$^{2)}$ Department of Physics and Astronomy,\\ 
University of Amsterdam,
Valckenierstraat 65, 1018 XE Amsterdam, The Netherlands 
\\ $^{3)}$Yerevan Physics Institute,
Alikhanian Brothers St. 2, Yerevan 375036, Armenia }
\date{\today}
\maketitle

\begin{abstract}
A model of quantum measurement is proposed, which aims to describe
statistical mechanical aspects of this phenomenon, starting from a 
purely Hamiltonian formulation. The macroscopic measurement apparatus 
is modeled as an ideal Bose gas, the order parameter of which, 
that is, the amplitude of the condensate, is the pointer variable. 
It is shown that properties of irreversibility and ergodicity breaking, 
which are inherent in the model apparatus, ensure the appearance of 
definite results of the measurement, and provide a dynamical realization 
of wave-function reduction or collapse. The measurement process takes 
place in two steps: First, the reduction of the state of the tested
system occurs over a time of order $\hbar/(TN^{1/4})$, where $T$ is
the temperature of the apparatus, and $N$ is the number of its degrees 
of freedom. This decoherence process is governed by the apparatus-system
interaction. During the second step classical correlations are
established between the apparatus and the tested system over the much 
longer time-scale of equilibration of the apparatus. The influence of
the parameters of the model on non-ideality of the measurement is 
discussed. Schr\"{o}dinger kittens, EPR setups and information
transfer are analyzed.
\end{abstract}


\renewcommand{\thesection}{\arabic{section}}
\section{ Introduction }
\setcounter{equation}{0}\setcounter{figure}{0} 
\renewcommand{\thesection}{\arabic{section}.}

Understanding the specific features of quantum measurements has been
a long-standing question 
\cite{bohr,wigner,landau,wh,vN,balian,bv,vK}. Discussions on 
this subject and its interpretations started in the early days of the quantum 
theory \cite{bohr,wigner,landau,wh,vN}. Physical and philosophical reflections on 
the problem and consideration of its different aspects became a source 
for deep conclusions about the quantum world and its classical counterpart 
\cite{wigner,wh}. Also the current activity in this field 
\cite{balian,bv,vK,rev,namiki,omnes} clearly displays its complexity 
and multifariousness, and witnesses that its final settling is not yet close.

The main purpose of this paper is to exhibit the complete solution of a
model for a measurement process, which will show that two main paradigms of
modern statistical physics, irreversibility and ergodicity breaking 
in phase transitions, are crucially relevant for the quantum
measurement problem. Together with some other conditions they provide all
necessary ingredients for the realization of ideal quantum measurements.
No additional postulates should be posed, since standard quantum 
statistical physics completely suffices for the self-consistent explanation
of this phenomenon.

We will start with a discussion on the problem of quantum measurement 
and the main steps which were made in its interpretation and 
understanding.

\subsection{General measurement }

Let us recall the general requirements that a measurement
should satisfy. The quantity to be measured in the 
system $\RS$ under study is represented by a hermitean operator 
$x$ with eigenvalues $x_k$. The so-called pointer variable $X$ in the
measuring apparatus $\RA$ may take values $X_k$ in correspondence with 
$x_k$. The observation of $X_k$ provides statistical information about
the state of the system. Moreover, some measurements termed as ideal
\cite{bv} can be used as filters to prepare the system in a state where
$x$ takes a well-defined value $x_k$. 

More precisely, we denote by $\rho $ the density operator of the system 
under study, by $R$ that of the apparatus, and by ${\cal R}$ the global
density operator. To include the possibility of the most
general descriptions of the system and the apparatus, we analyze 
the situation in terms of density operators rather than pure states.
For simplicity we will operate in this section with discrete spectra.
At the initial time $t=0$ just before measurement, 
the system is in an {\it unknown} state $\rho (0)$, the apparatus 
is in a fixed state $R(0)$ and these states are uncorrelated, 
so that the complete state is described by:
\BEA
{\cal R}(0)=\rho (0)\otimes R(0).
\label{1.1}
\EEA
The uncorrelated character of this state simply reflects the fact that
for $t<0$ the tested system and the apparatus were not interacting.
An evolution operator ${\cal T}$ transforms this initial density operator
into the overall density operator ${\cal R}(\theta )$ at the final time
$\theta $ of the measurement, which should have the form
\BEA
{\cal T}[\rho (0)\otimes {\cal R}(0)] 
\equiv {\cal R}(\theta )\simeq \sum _kp_k{\cal R}_k.
\label{1.4}
\EEA
Here ${\cal T}$ does not depend on the initial state of the system, 
which is arbitrary and unknown, though it may depend on the initial state 
of the apparatus $R(0)$. By $\simeq$ we mean ``as precise as desired 
provided that the parameters of the apparatus are tuned suitably''.

In a precise and unbiased measurement the final possible states ${\cal R}_k$
entering Eq.~(\ref{1.4}) should constitute an orthogonal set,
\BEA
{\rm tr}({\cal R}_k{\cal R}_l)\simeq\delta _{kl},
\label{1.01}
\EEA
so as to ensure that they represent {\it exclusive} events to which 
ordinary (non-quantum) probabilities can be assigned. They may be
distinguished from one another by observing the pointer variable $X$,
which takes in each ${\cal R}_k$ the value $X_k$ with a negligible statistical 
fluctuation, namely
\BEA
{\rm tr}({\cal R}_kX)=X_k,\qquad {\rm tr}({\cal R}_kX^2)\simeq X^2_k.
\label{1.2}
\EEA
Hereafter ${\rm tr}_\RA$ and ${\rm tr}_\RS$ indicate traces in the subspaces
of the apparatus and the system respectively, while ${\rm tr}$ is reserved 
for the complete trace.

Each $X_k$ occurs in Eq.~(\ref{1.4}) with a probability $p_k$, 
which is determined by the initial state of the system in the form
\BEA
p_k={\rm tr}_\RS (\rho (0)\Pi _k),
\label{1.3}
\EEA
where $\Pi _k=|x_k\rangle\langle x_k|$ denotes the projection operator
onto the eigenvalue $x_k$ of $x$ in the Hilbert space of the system.
Although the states $\R _k$ may depend on the initial state $\rho (0)$ 
of the particle, the quantities 
$X_k$ do not depend on $\rho (0)$ for precise and unbiased measurements, 
but are determined by the structure of the apparatus only. On the other hand,
the probabilities $p_k$ are determined by the initial state $\rho (0)$ only.

The fact that the process which leads from ${\cal R}(0)$ to ${\cal R}(\theta 
)$ describes a measurement is reflected in a specific feature. 
Due to the exclusiveness property (\ref{1.01}), Eq.~(\ref{1.4})
represents the occurrence of a {\it classical} random quantity $k$, with
probability $p_k$. It expresses that after the observation has taken place, 
the overall system is left in a state ${\cal R}_k$ with probability $p_k$. 
Indeed, the density operator $\R (\th )$ describes a statistical 
{\it ensemble of measurements} all performed under similar conditions 
rather than an individual experiment.
Since the variable $X$ has a definite value in each
state ${\cal R}_k$, one can count the frequencies of the different $X_k$'s 
and thus recover the unknown probability distribution $p_k$.
It is the specific type of correlation 
exhibited for each $k$ in Eq.~(\ref{1.4}),
between the properties (\ref{1.2}) pertaining to the apparatus and
the expression (\ref{1.3}) for the probabilities $p_k$ in terms of 
$\rho (0)$, which allows us to gain information about the tested system.

The crucial problem of quantum measurement is therefore to explain how
the evolution process of the coupled state $\R (t)$ from 
$\R (0)$ to $\R (\theta )$ can produce a transformation ${\cal T}$ which
ensures Eq.~(\ref{1.4}).
The subsequent process of {\it observation} then merely amounts to the 
{\it selection} of a single term of Eq.~(\ref{1.4}) characterized by the 
value $X_k$ of the pointer. This last step is by no means different 
from the analogous process in the {\it classical probability} theory 
\cite{balian,vK}, as was stressed recently by van Kampen and one of us. 
It should be stressed additionally that observation and selection refer to the
specific type of measurement process which has been previously performed.

\subsection{Ideal measurement}

The above measurement scheme is rather general, and in particular describes 
situations where the system itself does not have a definite state after
measurement (e.g., a photon is destroyed when detected by a photo-multiplier).

A theoretically and practically important class of measurements are the ideal 
ones, which leave the tested system as weakly perturbed as possible after
selection of the result.
In an ideal measurement, the possible final states $\R _k$ factorize as
\BEA
{\cal R}_k(\theta ) = \rho _k\otimes R_k,
\label{1.6}
\EEA
where $\rho _k$ is expressed in terms of the initial state 
$\rho (0) $ of the system as
\BEA
\rho _k =\frac{1}{p_k}\Pi _k \rho (0)\Pi _k,
\label{1.5}
\EEA
and where the possible final states of the apparatus, characterized by the
value $X_k$ of the pointer variable as in Eq.~(\ref{1.2}), are orthogonal
as in Eq.~(\ref{1.01}), 
\BEA
{\rm tr}( R_k R_l)\simeq\delta _{kl},
\label{bars}
\EEA
and are independent of the initial state $\R (0)$. Notice that the condition
(\ref{bars}) is stronger than that given by Eq.~(\ref{1.01}).
Later we will connect it with {\it robustness} of the apparatus as an
information-storing device.

Thus, after observation of the apparatus, and sorting of a given outcome 
$X_k$, the system is {\it prepared} in the state (\ref{1.5}) by means of 
the so-called ``reduction of the wave-packet'' or 
``collapse of the wavefunction''. 
Apart from bringing the system into an eigenstate of the observable $x$
associated with the eigenvalue $x_k$, the projection (\ref{1.5}) does not
affect its other degrees of freedom.

A complete measurement theory should provide conditions under which an 
apparatus interacting with the tested system brings it into one among
the reduced states $\rho _k$.

\subsection{The standard approach}

When they deal with quantum measurements, textbooks usually justify 
the above properties by relying on the conventional arguments initiated
by Bohr \cite{bohr,landau,rosen}. Somewhat qualitative and incomplete,
this general line of reasoning became known as the 
Copenhagen interpretation. Different, though closely connected
versions were summarized by Rosenfeld \cite{rosen} 
and more recently by van Kampen \cite{vK} among others.
The first precise discussion on the measurement problem was given by von 
Neumann \cite{vN}, who clarified
the issues, but was led to consider the properties of quantum measurements,
in particular the reduction (\ref{1.5}) as a {\it postulate}, which
complements the standard principles of quantum mechanics. 
This additional postulate is, however, not needed as one can
show, using consistency arguments, that the reduction of the wave packet can
be derived from the standard principles and from natural properties
attributed to measurements, in particular, repeatability \cite{balian}.

Nevertheless, a complete understanding of a quantum measurement requires
its analysis as a {\it dynamical process}.  A crucial point to be explained 
is the nonexistence in the final state (\ref{1.4}) of so-called
Schr\"odinger cat terms with respect to the index $k$. 
For simplicity let us specialize on an ideal measurement.
Starting from any initial state of the system, the final density operator
${\cal R}(\theta )$ should commute with both the measured observable $x$ 
of the system and the pointer observable $X$ of the apparatus. This property
ensures that the result of the measurement process can be described in the
language of classical probability theory. Otherwise, if ${\cal R}(\theta )$ 
did include off-diagonal contributions in $k$, the standard interpretation
of the measurement process would fail.

The Copenhagen school of thought overcomes this difficulty by saying that any 
physically acceptable apparatus has to be a classical system \cite{bohr,landau}.
Thus it cannot exists in a state with superpositions.
More refined Copenhagen-like approaches \cite{rosen,dan,vK} state that
the apparatus is a macroscopic system, and therefore coherent 
superpositions may arise, but cannot be detected at least when measuring
certain observables.
This situation was illustrated by an exactly solvable model \cite{hepp}
(see also \cite{cini} in this context),
where a class of observables was proposed, which are indeed 
non-sensitive to superpositions. Nevertheless, a small modification of this
class allows to produce superpositions \cite{bell}.

In the von Neumann-Wigner approach \cite{wigner,vN} the
states of the apparatus are pure, as $R(0)= |\Psi \rangle\langle\Psi |$, and
$R_k=|\Psi _k\rangle\langle\Psi _k |$, and
the evolution is characterized by the mapping 
\BEA
|\psi _k\rangle|\Psi \rangle\mapsto
|\psi _k\rangle|\Psi _k\rangle
\label{1.7}
\EEA
between initial and final states of the compound system.
If this evolution is applied to a coherent initial state of the
tested system such as
\BEA
\rho (0) = \left (
\sum _k \alpha _k|\psi _k\rangle
\right )\left (
\sum _{k'} \alpha ^*_{k'}\langle\psi _{k'}|
\right ),
\label{1.8}
\EEA
then one finally gets the following state
$\tilde{{\cal R}}(\theta )$ associated with $\rho (0)$:
\BEA
\tilde{{\cal R}}(\theta )=\sum_{k,k'}\alpha _k|\psi _k\rangle|\Psi _k\rangle ~
\alpha ^*_{k'}\langle\psi _{k'}|\langle\Psi _{k'}|,
\label{1.9}
\EEA
whereas the desired state ${\cal R}(\theta )$ 
in Eq.~(\ref{1.4}) involves only the diagonal elements
of $\tilde{{\cal R}}(\theta )$:
\BEA
{\cal R}(\theta )=\sum_{k}|\alpha _k|^2|\psi _k\rangle|\Psi _k\rangle
\langle\psi _{k}|\langle\Psi _{k}|.
\label{1.10}
\EEA
Partial traces of $\tilde{{\cal R}}$ and ${\cal R}$ over either the
apparatus or the system subspace are equal, but we should explain why the
off-diagonal terms of $\tilde{{\cal R}}$ are never seen.
Indeed, one has for the partial density matrix of the particle
\BEA
\rho (\th ) = {\rm tr}_{\rm A}\tilde{{\cal R}}(\th )
= {\rm tr}_{\rm A}{\cal R}(\th )=
\sum _k |\alpha _k|^2\, |\psi _k\rangle\langle\psi _k|. 
\EEA
However, the situation 
described by $\tilde{{\cal R}}(\th )$ does not correspond to 
any measurement, since the apparatus and the tested system cannot be in 
definite states with definite probabilities.
Actually, since with modern experimental equipments one is able to detect
mesoscopic and even {\it macroscopic} superpositions
(see e.g., \cite{nature1,nature2} where recent results are
reported in the context of charge and flux macroscopic 
superpositions in Josephson junctions), 
the question remains open and can be formulated in the following form: 
What are the concrete properties of a system, that make it usable as 
a measuring apparatus characterized by 
the collapsed state Eqs.~(\ref{1.1}--\ref{1.5}) ?

\subsection{Irreversibility}

An important requirement for the realization of Eq.~(\ref{1.1}--\ref{1.5})
is the existence of {\it irreversibility}.
Whereas (\ref{1.9}) is a pure state, we expect the pointer variable 
to take some well-defined value $X_k$ with the classical probability
$p_k$, and this implies the final state to be the mixture (\ref{1.10}).
In other words, the von Neumann entropy $S_{{\rm vN}}(\R (\theta))
=-{\rm tr}\R (\theta)\ln \R (\theta)$ of the state $\R (\theta)$ in
Eq.~(\ref{1.10}) is positive, while $S_{{\rm vN}}(\tilde{\R} (\theta))=0$.
More generally the von Neumann entropy of (\ref{1.4}) can be
shown to be different from that of (\ref{1.1}), due to the elimination 
of the coherent terms in the final state. This implies a {\it loss
of information} about the coherence, a loss which is required to
ensure the classical
interpretation of the measurement and the reduction of the wave packet.

A measurement process should therefore be analyzed in the same way as 
an irreversible process in quantum statistical mechanics, a second
reason for using density operators. The size of the apparatus should be 
sufficiently large, so that irreversibility and relaxation emerge 
from the microscopic reversible evolution. Otherwise measurements 
could not be ideal. 
The fact that the condition (\ref{1.4}) cannot be realized with
a unitary transformation from any initial state (\ref{1.1}) is nowadays 
well-established with different degrees of generalization \cite{wigner,abner}.
To understand this in simple terms, let us write down the ideal measurement
transformation for two different initial states $\rho ^{(s)}(0)$, 
$s=1,2$ of the system:
\BEA
{\cal T}[\rho ^{(s)}(0)\otimes R(0)] =\R^{(s)}(\theta ) 
\simeq \sum _kp^{(s)}_k \rho _k\otimes R_k,
\label{bars1}
\EEA
and assume for simplicity that the spectrum of $x$ is non-degenerate:
$\Pi _k=|x_k\rangle\langle x_k|$. Then unitarity of ${\cal T}$ 
requires ${\rm tr}_{\RS}[\rho ^{(1)}(0)\rho ^{(2)}(0)]=
{\rm tr}[\R^{(1)}(\theta ) \R^{(2)}(\theta )]$,
and conditions (\ref{1.5}, \ref{bars}) imply:
\BEA
{\rm tr}_{\RS}~\rho ^{(1)}(0)\rho ^{(2)}(0) \equiv
\sum _{k,l} \langle x_k|\rho ^{(1)}(0)|x_l\rangle
\langle x_l|\rho ^{(2)}(0)|x_k\rangle \simeq &&
\sum _{k} \langle x_k|\rho ^{(1)}(0)|x_k\rangle
\langle x_k|\rho ^{(2)}(0)|x_k\rangle \EEA
and hence
\BEA
\qquad &&\sum _{k\not =l} \langle x_k|\rho ^{(1)}(0)|x_l\rangle
\langle x_l|\rho ^{(2)}(0)|x_k\rangle \simeq 0,
\EEA
which cannot be true for arbitrary $\rho ^{(1)}(0)$ and $\rho ^{(2)}(0)$.
Thus, the unitarity of ${\cal T}$ has to be gotten rid of.
Actually, the situation is the same as in any relaxation process:
The overall system is isolated, and the evolution of $\R (t)$ is
in principle governed by Hamiltonian dynamics, but on suitable time-scales
irreversibility occurs owing to the presence of a large number of degrees
of freedom, which act as an external bath. Statistical physics is needed
to explain this behavior, in which microscopic reversible equations
of motion result in macroscopic irreversible ones.
Just the same approach was followed recently by two of us, to discover
that the standard issue of Brownian motion leads to incompatibilities
with thermodynamics in the regime of quantum entanglement~\cite{ANquant}.

\subsection{Ergodic and decoherence approaches to quantum measurements}

A first application of statistical physics in support
of the Copenhagen interpretation was 
given by Daneri, Longier and Prosperi \cite{dan,rosen}.
Somewhat related (but not equivalent) approaches are 
reviewed in Ref.~\cite{namiki}. After considering the measuring apparatus 
and measured system together as an
 {\it isolated} system, one attributes the absence 
of macroscopic superpositions to inevitable statistical uncertainties, which 
are present in macroscopic bodies. Mathematically this is reflected in 
different kinds of ergodic assumptions, which are reasonably creditable for 
those systems. However, these approaches have several drawbacks, which, 
in particular, originate from the fact that they do not provide dynamical 
mechanisms for the realization of quantum measurements. Extensive criticism 
on them can be found in \cite{rev}.

There is another, nowadays not less influential, school of thought which 
we shall follow. It attempts to handle the problem also involving certain
arguments from statistical physics \cite{zeh,zurek,rev}, \cite{omnes}.
In this {\it decoherence} approach the loss of coherence is viewed as a 
process established by an external environment, which is generally
understood as a collection of uncontrollable and unobservable degrees
of freedom. A decoherence process suppresses superpositions of some special 
states, which are determined by the interaction between the environment
and the system. However, in an ideal quantum measurement,
the coherence associated with both the observable $x$ of the tested system,
and the observable $X$ of the apparatus should disappear 
{\it independently} of
the concrete form of environment-system interaction, whereas the other 
coherences which exists in $\rho (0)$ should remain present in the
reduced states $\rho _k$ defined by Eq.~(\ref{1.5}). 
The type of decoherence occurring in quantum measurements is thus 
very special and we shall relate its features to the apparatus-system
interaction rather than to an environment-system interaction.

\subsection{Requirements on quantum measurement models}

Keeping in mind both  the successes and shortcomings of the various 
existing ideas,
we tackle in this paper the quantum measurement problem by investigating
a specific model. The coupled evolution
of the system and apparatus is treated as a dynamical process of 
quantum statistical mechanics. By deriving an explicit solution we wish 
to show how the various features of an ideal measurement, expressed by 
Eqs.~(\ref{1.1}--\ref{1.5}), can emerge from the microscopic
dynamics generated by the Hamiltonian of our model.

When choosing the model we have been guided by various conditions that an 
apparatus should satisfy:

1) It should have a degree of freedom $X$ which may relax towards definite
values $X_k$.

2) It should be macroscopic so as to ensure an irreversible relaxation.

3) This relaxation should be selectively triggered by the interaction of 
$X$ with the variable $x$ of the measured system. 

4) The various values 
$X_k$ should a priori be equally probable, so as to avoid any bias produced
by the apparatus. Thus, the various final states $R_k$, characterized by 
the value of $X_k$, should have the same entropy. 

5) The apparatus should be a stable and robust information storing device, 
which implies that the states $R_k$ are nearly in equilibrium and that after 
the measurement has been completed, $X$ is a nearly conserved collective 
variable. 

These properties suggest
to take for the apparatus a suitably chosen macroscopic system which is
able to undergo a phase transition, with $X_k$ as an {\it order parameter}.
Indeed, a phase transition is a macroscopic process with robust and stable
(or at least metastable)
outcomes. Notice that the existence of an order parameter implies ergodicity 
breaking in contrast to the purely ergodic view at measurement \cite{dan}.

6) The measured quantity $x$ should be coupled to the order parameter, 
and the apparatus should {\it amplify} this signal received during its 
interaction with the system. This is achieved by noting that the value 
of an order parameter can be controlled by an infinitesimally small source. 
The microscopic variable will play the role of such a source, which controls
$X$ but otherwise does not affect the apparatus. 

7) The relaxation of the
order parameter is ensured by its coupling with other degrees of freedom
of the apparatus, referred to as a thermal bath.
It is this coupling which, together with the thermodynamical limit 
for the apparatus, will ensure the specific type of relaxation
discussed above. 

We shall work out a model, as simple as possible, subject to the above requirements.
The tested system is a one-dimensional particle, and the quantity to be 
measured is its position. The apparatus is a non-interacting Bose gas, which
has an easily tractable phase transition. The variable $X$ is the amplitude
of the condensate. This situation will be shown to be  
generalizable for an arbitrary tested system and an arbitrary 
measured observable (Appendix A).
Although specific and not realistic 
the model is thus suggestive for more general measurements.
The possibility of tuning the parameters will help us to find the limit 
in which the measurement is ideal and to explore some imperfections of 
the measurement.

This paper is organized as follows. In section 2 we present the model
and discuss its equations of motion. Limits which are especially relevant
for the quantum measurement problem, as well as exact solutions of
the equations
of motion in the Schr\"odinger and Heisenberg pictures are considered in
section 3. In section 4 we show that the present model realizes the
conditions of ideal measurements discussed above. There we also 
consider characteristic times of this realization and discuss imperfections
which arise due to an incomplete thermodynamical limit for the apparatus. 
Our conclusions are presented in the last section. 
Several technical questions are considered in Appendices A, B, C and D.

\renewcommand{\thesection}{\arabic{section}}
\section{The model}
\setcounter{equation}{0}\setcounter{figure}{0} 
\renewcommand{\thesection}{\arabic{section}.}

\label{langevin}

\subsection{The apparatus and its bath}

As a model for our apparatus we will choose a system of $N$ non-interacting
bosons in a three-dimensional cubic box with volume 
$V$ and periodic boundary conditions. Its  Hamiltonian reads
\BEA
\label{HA}
H_{\rm A}=\sum_i\varepsilon _i a^{\dagger}_i a_i,
\EEA
where $a_i^{\dagger}$, $a_i$ are the
creation and annihilation operators of each single-boson state, and
$\varepsilon _i$ is its energy, given in terms of its wave vector $k$ 
and of the boson mass $M$ by:
\BEA
\label{spec}
\varepsilon _i = \frac{\hbar ^2k^2}{2M}.
\EEA
Notice that $\varepsilon_0=0$.
This apparatus is an open system, namely it interacts {\it weakly} 
with a large external environment. If there were no 
other interactions, the Bose gas would relax with time towards the Gibbs
distribution:
\begin{equation}
\label{G} \rho _{\RA}=\frac{1}{Z}~\exp (-\beta H_\RA +\beta \mu  
\N),\qquad
Z={\rm tr}\exp (-\beta H_\RA +\beta \mu \N),
\end{equation}
where $T=1/\beta$ is the temperature and $\mu $ is the chemical potential.
Both of these quantities are imposed by the environment. Finally $\N$
is the number operator
\BEQ \N=\sum_{i} a^\dagger_ia_i.
\EEQ

In a realistic apparatus what we call the ``external'' environment is 
actually constituted by a large number of degrees of freedom which
are part of the apparatus itself. Here we treat it as a separate thermal
bath, which can exchange energy and particles with the Bose gas. 
We can imagine this bath itself as a Bose gas which is much larger than
the apparatus, so that its intensive variables remain fixed once and forever.
The bath is characterized by its temperature $T$, its chemical 
potential $\mu $, and by a quantum coupling which induces 
a relaxation time $\gamma ^{-1}$ to the Bose gas.
This situation corresponds to the grand canonical ensemble for the
apparatus although the overall apparatus-bath system is isolated. 
Whereas the
value of $T$ is fixed by the overall energy, the value of $\mu $
is determined by the overall boson number. We shall focus below
on a condensed gas, with $|\mu |/T$ small as $1/\sqrt N .$ The
tuning of $\mu $ is then achieved through a control of the
overall particle number.

As a system examined by means of the apparatus, we take a particle
living in one-dimensional space, with mass $m$,
an external potential $V(x)$ and Hamiltonian
\BEA
H_{\rm S} = \frac{p^2}{2m}+V(x).
\EEA
The measured quantity is the position $x$, which is coupled to the Bose
gas through the interaction
\BEA
H_{\rm I} = -g~x~X,
\qquad 
X=\sqrt{\frac{\hbar}{2}}~(a_0^{\dagger}+a_0),
\label{kamaz}
\EEA
where $g$ is the coupling constant.
The quantity $X$ attached to the apparatus will be our pointer variable.
It is macroscopically accessible through observation of the density of 
the Bose gas provided Bose condensation takes place, which requires low
temperature and sufficiently small $|\mu |$. In this case we expect the 
signal $x$ to be amplified, because $X^2$ will be extensive.
On the other hand, a tested system
interacting with all levels of the apparatus would certainly be less 
interesting, since the threshold of the influence
on the apparatus would be diminished. We have chosen the simple
interaction (\ref{kamaz}) for theoretical purposes, although it
is not realistic since it can change the number of bosons.

Altogether the Hamiltonian of the apparatus and the system during the
measurement reads:
\BEA
H=H_{\rm A}-\mu {\cal N}+H_{\rm S}+H_{\rm I},
\label{kamaz11}
\EEA
to which we should add the interaction of the Bose gas with the bath,
see Appendix D.
Since the number of bosons may change through exchanges with the
bath and the observed system, we have included for convenience in Eq.
(\ref{kamaz11}) the contribution from the chemical potential.
Later on we will show how to generalize this situation to an arbitrary 
tested system and measured observable, keeping the same measurement 
apparatus.

\subsection{Bose condensation}
\label{bose}

As well known the three dimensional ideal Bose gas in
equilibrium undergoes a condensational phase transition 
at sufficiently large density or small chemical potential. Let us 
briefly recall this phenomenon, since in our setup this is 
a crucial property of the apparatus as an information-storing device. 
We consider here the Bose gas submitted to an external 
constant field source $J$, which later on will be identified with 
the term $\sqrt{\hbar /2}gx$ in $H_{\rm I}$:
\BEA
H_{\rm B} =\sum_i\varepsilon _i a^{\dagger}_i a_i -J(a_0+a_0^{\dagger}).
\label{hatha}
\EEA

We work in the grand canonical ensemble. When there is no source term,
it is known that despite different magnitudes of fluctuations in the condensed 
phase, the grand canonical and canonical ensembles are equivalent for the
non-interacting Bose gas \cite{Balian}. 
However, the source term in Eq.~(\ref{hatha}) can have a macroscopic
effect only in the grand canonical ensemble. It controls the 
density of the condensate, which can vary owing to possible exchange with the 
bath. In the canonical ensemble where the overall density is
given, the density of the condensate would be 
practically insensitive to the presence of the source. In our present 
situation the {\it density},
and not only the {\it phase} of $\langle a_0\rangle$, appears as an order parameter 
controlled by the field $J$. This property is specific for our model of 
non-interacting bosons, and would be invalid for a realistic Bose condensate 
with interaction.

At the equilibrium state with temperature $T$ and (negative) chemical 
potential $\mu$ one diagonalizes the gibbsian density matrix by shifting 
the lowest energy operators as 
\BEA
a_0=\tilde{a}_0-\frac{J}{\mu}, 
\qquad a^{\dagger}_0=\tilde{a}^{\dagger}_0-\frac{J}{\mu}.
\EEA
This leads to
\BEA
\langle a_0\rangle =-\frac{J}{\mu}, 
\label{urumchi1}
\EEA
\BEA
\langle a^{\dagger}_0a_0\rangle =
\frac{1}{{\rm e}^{-\beta \mu} -1}
+\frac{J^2}{\mu ^2}.
\label{urumchi2}
\EEA
The averages concerning excited states will not change, since the field 
acts only on the lowest mode. For the total density of particles one gets
\BEA
\label{ra1}
\frac{N}{V}=\frac{1}{V}\sum _i\langle a_i^{\dagger}a_i\rangle 
=\frac{(2M)^{3/2}}{4\pi ^2\hbar ^3} \int _0^{\infty}\frac{\d \varepsilon ~ 
\sqrt{\varepsilon} }{\re^{\beta (\varepsilon -\mu)} -1}
+\frac{1}{V}\frac{1}{\re^{-\beta \mu} -1}+\frac{J^2}{V\mu ^2},
\EEA
Here we went to the thermodynamical limit, which, if $x$ and thus $J$
are well defined, means to make the following change:
\BEA
\label{lore}
{\rm tr}~(...)\mapsto \int \frac{\d ^{3} x~\d ^{3}k}{(2\pi )^3}~(...)
=V\int \frac{\d ^{3}k}{(2\pi )^3}~(...)
=V\frac{(2M)^{3/2}}{4\pi ^2\hbar ^3}
\int_0^{\infty}\d\varepsilon~\sqrt{\varepsilon}(...).
\EEA
We also separated out the contribution coming from the 
lowest state $\varepsilon =0$.

We wish to measure $x$, which, if $x$ is well defined, means that
we wish to find the value of 
$J=\sqrt{\hbar /2}gx$ through a macroscopic observation of the Bose gas.
This is feasible by observing the total density (\ref{ra1}), provided that 
the last term in this equation is finite in the thermodynamical limit, which
requires: $|\mu | \to 0$ when $N\to \infty$. 
In this case the particle density 
\BEA
\label{ramon}
\frac{N}{V}=\frac{N_{\rm n}}{V}+\frac{N_{\rm c}}{V}
\EEA
splits into a non-condensed part,
\BEA
\label{ramon-ramon}
\frac{N_{\rm n}}{V}=
\frac{(2M)^{3/2}}{4\pi ^2\hbar ^3} \int _0^{\infty}\frac{\d\varepsilon ~
\sqrt{\varepsilon}}{\re ^{\beta \varepsilon} -1}
=\frac{0.165869~ M^{3/2}}{\hbar ^3}~T^{3/2},
\EEA
and a condensed part
\BEA
\label{okun}
\frac{N_{\rm c}}{V}=\frac{T}{V|\mu |}+\frac{J^2}{V\mu ^2},
\EEA
which is our order parameter. Since $N_{\rm n}/V$ is a known function of
temperature, we can deduce $N_{\rm c}/V$ from the total density, provided
that this total density is significantly larger than the critical value (\ref{ramon-ramon}):
\BEA
\frac{N}{V}>\frac{N_{\rm c}}{V}.
\EEA

In order to use the Bose gas as a measurement device for 
$J$, we shall require Eq.~(\ref{okun}) to be dominated 
by its last term, which means that
\BEA
\label{mazarishari}
1\gg \frac{|\mu |}{T}\gg \frac{1}{N},
\EEA
and that 
\BEA
\label{mazarishari1}
\frac{|\mu |}{J}={\cal O}\left (\frac{1}{\sqrt{N}}\right ).
\EEA
Under such conditions, the scalar source term $x$, which will be replaced 
later on by the coordinate of the tested particle, can be deduced from
the density of the Bose gas and the characteristics of its bath:
\BEA
x^2 = \frac{2\mu ^2V}{\hbar g^2}\left (
\frac{N}{V}-\frac{N_{\rm n}}{V}\right ).
\EEA
The sign of $x$ results through Eq.~(\ref{urumchi1}) from the phase of the
condensate. Low temperatures will improve the efficiency of the measurement, 
since they provide small values for (\ref{ramon-ramon}) 
and thus increase the ratio $N/N_{\rm n}$.
The chemical potential
of the bath should be fixed at some small value when $N\to \infty$,
so as to ensure Eq.~(\ref{mazarishari}, \ref{mazarishari1}), and $g$ should 
be such that as $g^2x^2\hbar/\mu ^2$ is of order $N$.

The excited states of the Bose gas will not play a direct role in our model 
of measurement, since they are not coupled to the tested particle. 
Nevertheless, they contribute, together with the bath, to exchange bosons with 
the condensate, the density of which can thus be controlled by the chemical
potential as well as by the source $J$.

Notice also that, although the various macroscopic states characterized 
by different values of the order parameter $N_{\rm c}/V$ can be distinguished,
they appear on the same footing, because both their entropy and their energy
are the same in the thermodynamical limit. Indeed the
contribution of the condensate to the entropy is $\ln N_{\rm c},$ its
contribution to $\left\langle H_{\rm B}\right\rangle $ is $2\mu N_{\rm c}$
 where $|\mu|\ll T,$ so that both become negligible as 
$N\rightarrow \infty$. This was required to prevent the apparatus from having 
an intrinsic bias.

\subsection{Equations of motion}

\subsubsection{Dynamics of the apparatus in its bath}
Before we examine the equations of motion of the overall system
including the tested particle, the apparatus and the bath, we will investigate
in this subsection the situation without the tested particle.
At some
remote initial time $t=t_0$ the apparatus was in an
arbitrary {\it non-equilibrium} state. 
At that time it starts to interact with the bath, and 
the arising dynamics for the apparatus
will be described by means of a weak-coupling 
quantum Langevin equation \cite{gardiner} 
(see Appendix D for the derivation of this equation from the Heisenberg
equation associated with the apparatus-bath Hamiltonian):
\begin{eqnarray}
\label{vavilon100}
\dot{a}_i=\frac{\ri}{\hbar}[H_{\rm A}-\mu {\cal N},a_i(t)]-\gamma a_i+\sqrt{2\gamma }~b _i(t)
= - \ri (\omega _i +\alpha )a_i -\gamma a_i+\sqrt{2\gamma }~b _i(t),
\end{eqnarray}
where 
\BEA
\hbar \omega _i\equiv \varepsilon_i,\qquad \hbar \alpha \equiv -\mu.
\EEA
Notice that the chemical potential $\mu$ is negative, whereas
the parameter $\alpha$ is positive. The Langevin equations are written for 
the Heisenberg operators of the apparatus only, but the presence of the bath
is reflected through a friction term $-\gamma a_i$, and a random Gaussian 
force operator $b _i(t)$, which satisfies:
\BEA
\label{s1}
&&[b_i(t),b^{\dagger}_k(t')]=
\delta _{ik}\delta (t-t'),
\qquad [b_i(t),b _k(t')]
=[b^{\dagger} _i(t),b^{\dagger}_k(t')]=0,
\\
\label{s2}
&&\langle b^{\dagger}_i(t)b_k(t')\rangle
=\delta _{ik}\delta (t-t') n_i^{\rm eq} ,
\qquad \langle b_i(t)b_k(t')
\rangle=\langle b^{\dagger}_i(t)
b^{\dagger}_k(t')\rangle=0,
\\ \label{s3}
&&n_i^{\rm eq} =\frac{1}{{\rm e}^{\beta (\varepsilon _i-\mu )} -1}.
\EEA

The most important consequence of
Eqs.~(\ref{vavilon100}, \ref{s1}, \ref{s2}, \ref{s3}) 
is that they ensure relaxation with the characteristic 
time $1/\gamma$ of the apparatus towards the Gibbs distribution
(\ref{G}) with the temperature $T$ and chemical potential $\mu$ imposed by 
the bath. This can be seen from the following exact solution of 
Eq.~(\ref{vavilon100}):
\BEA
a_i(t)=
\re ^{-\gamma (t-t_0)-\ri (\omega _i+\alpha )(t-t_0)}a_i(t_0)
+\sqrt{2\gamma}\int _{0}^{t-t_0}\d s~\re ^{-\gamma s-\ri (\omega _i+\alpha )s}
~b_i(t-s).
\label{plumbum0}
\EEA
In particular, all possible moments $\langle a^{\dagger\, n}(t)a^m(t)\rangle$
calculated with Eq.~(\ref{plumbum0}) for 
$t-t_0\gg 1/\gamma$ are identical to those obtained through
the Gibbs distribution (\ref{G}). For example the average number of particles
$\langle a_i^{\dagger}(t)a_i(t)\rangle$ 
in the level $i$ evolves in time according
to
\BEA
\langle a_i^{\dagger}(t)a_i(t)\rangle\equiv n_i(t)= 
\re ^{-2\gamma (t-t_0)}n_i(t_0)+(1-\re ^{-2\gamma (t-t_0)})n_i^{\rm eq},
\label{batu0}
\EEA
which shows that $n_i(t)$ relaxes to its gibbsian stationary value
$n_i^{\rm eq}$ at the characteristic time $1/(2\gamma )$.

Since as shown by 
Eqs.~(\ref{kamaz}, \ref{kamaz11}) the tested particle interacts only
with the lowest level of the apparatus, the equations of motion for the excited
levels with $i\ge 1$ will be always given by Eq.~(\ref{plumbum0}). Therefore,
in the further discussion we will leave these excited levels aside.

\subsubsection{Equations of motion including the tested particle}

Let us now consider the situation with the tested system.
The interaction between the apparatus and the tested system is switched 
on at the initial time $t=0$. For $t\le 0$ the overall initial state
factorizes as in Eq.~(\ref{1.1}), where $\rho (0)$ is an arbitrary
state of the tested particle and $R(0)$ is the Gibbs distribution of
the apparatus given by Eq.~(\ref{G}). Indeed due to the assumed
condition $t-t_0\gg 1/\gamma$ the apparatus had enough time to relax
starting from any initial state at $t=t_0$. From now on we shall 
drop the index $0$ in $a_0=a$, $b_0=b$.
The Heisenberg--Langevin equation of motion for the lowest level of the
apparatus reads for $t>0$:
\begin{eqnarray}
\label{vavilon1}
\dot{a}=\frac{\ri}{\hbar}[H,a(t)]-\gamma a+\sqrt{2\gamma }~b (t)
= - \ri \alpha a +\frac{\ri}{\sqrt{2\hbar}}~ 
g~x(t)-\gamma a+\sqrt{2\gamma }~b (t).
\end{eqnarray}
This equation is solved exactly as
\BEA
a(t)=
\re ^{-\gamma t-\ri \alpha t}a(0)
+\frac{\ri ~ g}{\sqrt{2\hbar}}
\int _0^t\d s~\re ^{-\gamma s-\ri \alpha s}~x(t-s)
+\sqrt{2\gamma}\int _0^t\d s~\re ^{-\gamma s-\ri \alpha s}
~b(t-s).
\label{plumbum}
\EEA
For $\gamma t\gg 1$, and when $x$ is constant, Eq.~(\ref{plumbum}) expresses
that the Bose gas relaxes towards an equilibrium state, where 
the particle number in the condensate is given by 
Eq.~(\ref{okun}) with $J=\sqrt{\hbar /2}gx,$ as ensured by the second term
in the r.h.s. of Eq.~(\ref{plumbum}).
The average number of particles $n(t)=\langle a^{\dagger}(t)a(t) \rangle $
in the lowest state is evolving as
\BEA
n(t)= \re ^{-2\gamma t}n(0)
+\frac{g^2x^2}{2\hbar (\gamma ^2+\alpha ^2)}
\left (
1+\re ^{-2\gamma t}-2\re ^{-\gamma t}\cos \alpha t
\right )+(1-\re ^{-2\gamma t})n^{\rm eq}.
\label{batu}
\EEA
The second term in the r.h.s. of Eq.~(\ref{batu}) is the contribution 
supplied by the source, which shifts the condensate density. 

Although the evolution of the apparatus when there is no source leads
to a well-defined equilibrium state where $n(0)=T/ |\mu |$ is
large but not extensive as $V \rightarrow \infty ,$ the small interaction with the tested
particle is sufficient to change macroscopically $n$ at times $t\gg
1/(2\gamma)$ if $|\mu |=\hbar\alpha$ is sufficiently small. This means
that the
apparatus together with its bath constitute a system which is 
{\it nearly non-ergodic} when Bose condensation sets in.

The Heisenberg dynamics of the particle reads:
\BEA
\dot{F}=\frac{\ri}{\hbar}[H_{\rm S},F(t)]
-\frac{\ri ~g}{\sqrt{2\hbar}}[x(t),F(t)]
(a(t)+a^{\dagger}(t))
\label{vavilon2}
\EEA
for any operator $F$. We find in particular
\begin{eqnarray}
\label{vavilon6}
&& \dot{x}= \frac{1}{m}p,
\\
&&\label{vavilon7}
\dot{p}= - V'(x) +
\int _{0}^{t}\d s \chi (t-s)x(s)+\eta (t),
\EEA
where
\BEA 
\label{etat=}
&&\eta (t)=\eta _0(t)+\eta _1(t), \\
&&\eta _0(t)=
\sqrt{\hbar \gamma}~
g\int _0^t\d s~\re ^{-\gamma s}(b ^{\dagger}(t-s)
\re ^{\ri\alpha s}
+b(t-s)\re ^{-\ri\alpha s}),\\
&&\eta _1(t)=\sqrt{\frac{\hbar}{2}}~
g\re ^{-\gamma t}(a^{\dagger}(0)\re ^{\ri\alpha t}
+a(0)\re ^{-\ri\alpha t}),
\EEA
\BEQ \label{hh=}
\chi (t)=g^2~
\re ^{-\gamma t}\sin \alpha t .
\EEQ
The interaction of the particle with the Bose gas produce a force, 
which has a random part. The randomness of the noise $\eta (t)$ 
arises from two independent reasons: the statistical (uncertain) character
of the initial state of the apparatus, which gives the contribution 
$\eta _1(t)$, and the random character of $b$, $b^{\dagger}$, 
which occurs through $\eta _0(t)$. Recall that at $t=0$ the apparatus was
in equilibrium at temperature $T$ and chemical potential $\mu$,
\BEA
\langle a(0)a^{\dagger}(0)\rangle = 1+\frac{1}{\re ^{-\beta \mu}-1},
\qquad \langle a^{\dagger}(0)a(0)\rangle = \frac{1}{\re ^{-\beta \mu}-1}.
\EEA
Since $b(t)$, $b^{\dagger}(t)$ are themselves gaussian, 
$\eta (t)$ will be gaussian as well, with the noise autocorrelation:
\BEA
\label{kalgata0}
&&K(t,t')=\langle \eta (t);\eta (t')\rangle =K_0(t,t')+K_1(t,t'),\\
\label{kalgata1}
&&K_0(t,t')
=\langle \eta _0(t);\eta _0(t')\rangle =
\frac{g^2\hbar}{2}\cos [\alpha (t-t')]~
{\rm coth}\frac{\hbar\alpha  }{2T}~\left (\re ^{-\gamma |t-t'|}-
\re ^{-\gamma (t+t')}\right ),\\
\label{kalgata2}
&&K_1(t,t')=\langle \eta _1(t);\eta _1(t')\rangle 
=\frac{g^2\hbar}{2}\cos [\alpha (t-t')]
~ {\rm coth}\frac{\hbar\alpha  }{2T}~
\re ^{-\gamma (t+t')},
\EEA
where we define for any operators $A$, $B$:
\BEA
\langle A;B\rangle \equiv
\frac{1}{2}\langle AB+ BA\rangle .
\EEA
Notice that $K(t,t')$ is time-translation invariant, although its separate 
parts are not. It is seen that for $t+t'\gg 1/\gamma $, which
corresponds to the stationary apparatus,
only $K_0(t-t')$ persists.
The most important effect of the bath on the dynamics of the tested 
particle is the appearance of a new characteristic correlation
time $1/\gamma $ in addition to the time-scale set up by the maximal
frequency of the apparatus and to the universal quantum correlation
time $\hbar /T$.

\subsubsection{Brownian motion of the tested particle}

In order to compare Eqs.~(\ref{vavilon6} -- \ref{hh=}) with the standard quantum Brownian 
motion approach, one integrates Eq.~(\ref{vavilon7}) by parts. 
This yields an equation, \cite{gardiner},
\BEA
\dot{p}= - V'(x) -\frac{1}{m}~
\int _{0}^{t}\d s \tilde{\chi } (t-s)p(s)+\eta (t)
-\tilde{\chi }(t)x(0)+\tilde{\chi }(0)x(t),
\label{barro}
\EEA
which has the usual form in terms of a friction kernel
\BEA
\tilde{\chi }(t)=\re ^{-\gamma t}
\frac{\alpha \cos \alpha t+\gamma\sin \alpha t }{\alpha ^2+\gamma ^2},
\EEA
and of a noise $\eta (t)$. Notice, however, that
the friction kernel does not have a definite sign with $t$.
The last term in Eq.~(\ref{barro}) renormalizes the potential.
As far as one is interested in the state of the tested particle itself, 
the full noise acting on it is $\eta (t)$. However, for the global 
state of the particle and the apparatus $\eta _1(t)$ is a deterministic 
object, and only $\eta _0$ remains as noise.

\subsubsection{Validity of weak-coupling quantum Langevin equations}

When substituting the above white-noise quantum Langevin equations
for the actual interaction between the apparatus and the bath,
a crucial fact was that their coupling is weak, so that the damping time $1/\gamma$ is much larger than both
the (maximal) dynamical characteristic time $t_{\rm d}$ of the
apparatus, and the characteristic 
correlation time of the bath $\hbar /T$.
Under these conditions it is possible, as shown in Appendix D, 
to introduce an effective quantum
noise $b (t)$ with white spectrum satisfying 
Eqs.~(\ref{s1}-\ref{s3}),
and get the Gibbs distribution as the result of relaxation. 
The time $t_{\rm d}$ is expressed as
\BEA
\label{imp}
t_{\rm d}\sim \frac{1}{\alpha} = \frac{\hbar}{|\mu |},
\EEA
as seen from the free part of Eq.~(\ref{vavilon1}).
Altogether the parameters of the bath should satisfy:
\BEA
\label{und}
\alpha \gg \gamma  \qquad {\rm or }\qquad
T\gg |\mu |\gg \hbar\gamma ,
\EEA
where we have taken into account the upper bound (\ref{mazarishari}) on
$|\mu |$.
Notice that excited levels of the Bose gas has a lower dynamical time 
\BEA
t^{(i)}_{\rm d}\sim  \frac{1}{\alpha +\omega _i}.
\EEA
It is expected that in a non-ideal Bose gas these characteristic times
will influence also the lowest mode, making its characteristic dynamical
time lower. 
The weak-coupling Langevin equation is discussed with more details in
Appendix D.

\renewcommand{\thesection}{\arabic{section}}
\section{DYNAMICS OF THE MEASUREMENT PROCESS}
\setcounter{equation}{0}\setcounter{figure}{0} 
\renewcommand{\thesection}{\arabic{section}.}

\subsection{Approximate conservation of the measured quantity.}

The above equations describe the joint evolution of the tested particle
and the Bose gas in the bath which ensures relaxation. This evolution will
describe a measurement of $x$, if the Bose condensate registers the statistical
distribution of $x$ in the initial state $\rho (0)$ under 
the conditions specified in the introduction.
In the present section we will take into account the following
condition for ideality of the measurement. 

If $x(t)$ did change with time during the process, the 
final state of the apparatus would be determined not only by 
the statistical distribution of
$x(0)$ but rather by the whole history of its change.
So to be sure that we indeed measure the quantity $x(0)$, we will require 
that the characteristic time for $x(t)$ to change appreciably from its 
initial value is much larger than the relaxation time $1/\gamma$
of the apparatus. 
This time is itself shorter than the duration $\theta$ of 
the measurement, since for $t=\th $ the apparatus should display
definite results through well-defined stationary states. 
Notice the conservation of $x(t)$ does
not mean that the state of the particle is not changing.
One only needs approximate conservation of the quantity
$x$. For the considered model this is realized if over
the time scale $\theta$ one can neglect the 
contribution of the kinetic energy $p^2/(2m)$ to the change of the
coordinate $x$. Obviously, this is realized for sufficiently large $m$. 
The more precise condition $m\gg g^2\th ^3$ will be found in Appendix C,
where we discuss the deviations from non-ideality which arise from the
motion of the particle during the measurement.
For other systems this conservation of the measured quantity may result
from a commutation relation: $[H_{\rm S},x]=0$, as stressed by
Wigner \cite{wigner} (see also \cite{yan} in this context). 
However, the actual mechanism is basically irrelevant provided 
that the time-scale for the variation of $x$ is much larger than the
duration of the measurement. 

Notice that taking the limit $m\rightarrow \infty $ as we do in
the bulk of this paper does not mean that we are dealing with a
classical particle. In fact, the initial density matrix $\langle
x'|\rho (0)|x'' \rangle $ of the particle is arbitrary, which
expresses that the quantity $x$ is subject to a fully quantum
probability distribution at the initial time owing to the
presence of off-diagonal elements. The large mass enters the
equation of motion (\ref{vavilon6}), and it only implies that the
diagonal elements $\langle x|\rho (t)|x\rangle $ remain
unchanged during the measurement.

\subsection{Wigner function and intermediate Wigner function}

\subsubsection{Wigner function}
For solving the dynamics in the Schr\"odinger picture,
we find it useful to employ the language of the Wigner function.
The description through Wigner functions is, of course, completely
equivalent to that through density matrices, but frequently, and in
particular for quantum systems with linear dynamical evolution equations, 
this allows to obtain and understand 
results in more easy way. This is because Wigner functions
allows to use the classical intuition as much as possible, which 
appears to be insightful for such systems.

We shall focus on the degrees of freedom $x$, $p$ of the tested particle and
$a$, $a^{\dagger}$ of the condensate, leaving aside the excited states of the 
Bose gas which relax independently.
To introduce a Wigner function not only for the particle but also
for the apparatus degrees of freedom, we represent the annihilation and
creation operators in the canonical way:
\BEA
&& a=\frac{1}{\sqrt{2\hbar}}(X+\ri P),\qquad 
a^{\dagger}=\frac{1}{\sqrt{2\hbar}}(X-\ri P),\\
&& X=\sqrt{\frac{\hbar}{2}}\left( a+a^{\dagger} \right ),
\qquad P=\ri\sqrt{\frac{\hbar}{2}}\left( a^{\dagger} -a\right ),
\\
&&[X,P]=\ri \hbar .
\EEA
Recall that the connection between a density matrix 
$\langle x'|\rho |x'' \rangle$ and 
the corresponding Wigner function $w (x,p)$ is given 
for each degree of freedom as
\BEA
\label{rhoW}
&&
\langle x+\frac{\xi}{2}|\rho |x-\frac{\xi}{2}\rangle =
\int \frac{\d p}{2\pi\hbar}~
\re^{\ri \xi p/\hbar} ~w (x,p),
\\
\label{Wrho}
&&w (x,p)=\int\d \xi ~\re^{-\ri \xi p/\hbar}
~\langle x+\frac{\xi}{2}|\rho |x-\frac{\xi}{2}
\rangle . 
\EEA
Notice that in the present paper the normalization of the Wigner function is
chosen as
\BEA
\int \frac{\d p~\d x}{2\pi\hbar}~w (x,p)=1,
\EEA
since the integration with this weight corresponds to a trace.
The Wigner function of the particle and the 
condensate together will be denoted by $\W (X,P,x,p)$; those of the
particle and condensate separately will be denoted as $w(x,p)$ and
$W(X,P)$, respectively. Obviously, one has
\BEA
\int \frac{\d X~\d P}{2\pi \hbar }~\W (X,P,x,p)=w(x,p),\qquad
\int \frac{\d x~\d p}{2\pi \hbar }~\W (X,P,x,p)=W(X,P).
\EEA

The Wigner function at time $t$ in the Schr\"odinger picture
can readily be represented in terms of the Heisenberg operators. 
For example, for the tested particle the corresponding formula reads
\BEA
\label{kismet}
&&w(x,p;t)= \langle {\rm tr}\rho (0)~\hat{w}(x,p;t)
\rangle ,
\nonumber \\
&& \hat{w}(x,p;t)=\int \frac{\d a~\d b}{4\pi ^2}
\exp [-iax-ibp+iax(t)+ibp(t)],
\EEA
where $x(t)$, $p(t)$ are the Heisenberg operators of the particle,
$\rho (0)$ is its initial state, and the average is
taken with respect to the full noise $\eta=\eta_0+\eta_1 $.

\subsubsection{Intermediate Wigner function}
\label{rurk}

We shall also find it convenient to use yet another formulation,
where the degrees of freedom of the tested particle are left in the matrix 
representation, and the Wigner transformation (\ref{Wrho}) is taken only for
the apparatus degrees of freedom. We will call this object 
the intermediate Wigner function, and
denote it as ${\cal V}(X,P,x',x'')$, where $x'$, $x''$ denote 
the corresponding matrix elements in the $x$-representation,
\BEA
{\cal V}(X,P,x',x'')=
\int\d \xi ~\re^{-\ri \xi P/\hbar}
\langle X+\frac{\xi}{2},x'|\R |X-\frac{\xi}{2} ,x''\rangle =
\int \d p~\re ^{\ri p(x'-x'')/\hbar} {\cal W}(X,P,\frac{x'+x''}{2},p).
\label{kamaz1}
\EEA
Here again the excited states of the Bose gas are left aside.

\subsection{Exact solution of the equations of motion in the 
Heisenberg picture}

We consider a heavy ($m\to\infty$), free ($V(x)=0$) particle. 
The equations of motion for the apparatus
have already been solved as Eq.~(\ref{plumbum0}) for the excited
states and Eq.~(\ref{plumbum}) for the lowest state. 
The latter equation, in which $x(t-s)$ can be replaced by $x(0)$,
is written in terms of $X$, $P$ and $b=b_0$ as
\BEA
\label{tulukan1}
X(t) =&& X(0)~\re ^{-\gamma t}\cos\alpha t 
+ P(0)~\re ^{-\gamma t}\sin\alpha t
+\frac{x(0)g}{\gamma ^2+\alpha ^2}
[\alpha -(\gamma\sin\alpha t +\alpha \cos\alpha t)~\re ^{-\gamma t}]
\nonumber\\
+&&\sqrt{\hbar\gamma}\int_0^t\d s~ \re ^{-\gamma s}
\left [
\re ^{-\ri \alpha s}b(t-s) + \re ^{\ri \alpha s}b^{\dagger}(t-s)
\right ],\\
P(t) =&& P(0)~\re ^{-\gamma t}\cos\alpha t 
- X(0)~\re ^{-\gamma t}\sin\alpha t+\frac{x(0)g}{\gamma ^2+\alpha ^2}
[\gamma +(\alpha\sin\alpha t -\gamma \cos\alpha t)~\re ^{-\gamma t}]
\nonumber\\
+&&\ri \sqrt{\hbar\gamma}\int_0^t\d s~ \re ^{-\gamma s}
\left [
\re ^{\ri \alpha s}b^{\dagger}(t-s) - \re ^{-\ri \alpha s}b(t-s)
\right ].
\label{tulukan2}
\EEA

For the tested particle we can solve Eqs.~(\ref{vavilon6}, \ref{vavilon7}) 
when the operator $x(t)$ does not significantly change as
\BEA
\label{vavilon77}&& x(t)=x(0),\\
&&p(t)=p(0) +
x(0)\xi (t) +
\int _0^t \d u ~ \eta (u),
\label{hu}
\EEA
where
\BEA 
\xi (t)=\int _0^t\d u\int _0^u\d s~\chi (s)
=g^2~
\frac{(\gamma ^2 +\alpha ^2) \alpha t+(\gamma ^2-\alpha ^2)
\re^{-\gamma t}\sin \alpha t-2\alpha\gamma
(1-\re^{-\gamma t}\cos \alpha t) }
{(\gamma ^2 +\alpha ^2)^2},
\label{sarancha}
\EEA
and $\chi (t)$, $\eta (t)$ are defined by Eqs.~(\ref{etat=}~-~\ref{hh=}).
Hereafter the following formulas will be used:
\BEA
\int _0^t\d s\re^{-\gamma s}\cos \omega s=
\frac{ \gamma [1-\re^{-\gamma t}\cos \omega t] 
+ \omega \re^{-\gamma t}\sin\omega t         }
{    \gamma ^2+\omega ^2    },
\qquad 
\int _0^t\d s\re^{-\gamma s}\sin \omega s=
\frac{ \omega [1-\re^{-\gamma t}\cos \omega t] 
-\gamma \re^{-\gamma t}\sin\omega t         }
{    \gamma ^2+\omega ^2    }
\EEA

\subsection{
Exact solution of the equations of motion in the Schr\"odinger picture}

Since the above dynamical equations are linear, there is a direct
connection between the Heisenberg picture and the Schr\"odinger dynamics
in terms of the overall Wigner function for the tested particle and the
apparatus. The equation for the common Wigner function of the particle
and the lowest mode has the form
\BEA
{\cal W}(X,P,x,p;t)=\int \d \BX~\d \BP~\d \Bp 
~\Phi (X,P,x,p ;t~|~ \BX,\BP,x,\Bp ;0)
W(\BX ,\BP;0)w(x,\Bp ;0).
\label{mr}
\EEA
Here we denote by $\BX , \BP , \Bp$ the variables of the Wigner function 
at the initial time. The variable $x$ of the Wigner function remains 
unchanged
from the initial to the final state, since the Heisenberg operator 
$x(t)$ is conserved. In Eq.~(\ref{mr}) we used the fact that the 
initial Wigner function ${\cal W}$ is factorized into the partial 
Wigner function of the 
lowest level $W(0)$ and that of the tested particle $w(0)$:
\BEA
{\cal W}(X,P,x,p ;0)=W(X,P;0)w(x,p;0)=
w(x,p;0)~\frac{\hbar}{\lambda}
\exp \left [
-\frac{1}{2\lambda}
X^2
-\frac{1}{2\lambda}
P^2 \right ],
\label{kaban}
\EEA
where we took into account the fact that at the initial time $t=0$ the apparatus
already relaxed to the equilibrium Gibbs distribution under the influence 
of the bath.

To find the transition kernel $\Phi$, we notice that
the Heisenberg operators $X(t)$, $P(t)$ and $p(t)$ of the Bose gas and
the tested particle given by Eqs. (\ref{tulukan1} -- \ref{hu}) are
{\it linear} and involve only the {\it gaussian} noise $b,b^{\dagger}.$ 
We can then use the method of Eq.~(\ref{kismet}) for the whole system.
Recall that the moments of the Wigner function 
coincide with the corresponding {\it symmetrized} operator moments.
Hence, $\Phi $ has the gaussian form
\BEA
\Phi (X,P,x,p ;t~|~ X_0,P_0,x,p_0;0)=
\frac{\sqrt{{\rm det}~B}}{(2\pi)^{3/2}}
~\exp \left (
-\frac{1}{2} \sum _{i,k=1}^3B_{ik}L_iL_k\right ),
\label{kr}
\EEA
where $(L_1,L_2,L_3)$ is the following three-dimensional vector:
\BEA
\label{kamchadal0}
&&L_1=X-\langle X\rangle = X
-X_0~\re ^{-\gamma t}\cos\alpha t 
-P_0~\re ^{-\gamma t}\sin\alpha t
-\frac{xg}{\gamma ^2+\alpha ^2}
[\alpha -(\gamma\sin\alpha t +\alpha \cos\alpha t)~\re ^{-\gamma t}],
\\
&&L_2=P-\langle P\rangle =P
-P_0~\re ^{-\gamma t}\cos\alpha t 
+ X_0~\re ^{-\gamma t}\sin\alpha t  - \frac{xg}{\gamma ^2+\alpha ^2}
[\gamma +(\alpha\sin\alpha t -\gamma \cos\alpha t)~\re ^{-\gamma t}],
\\
&&L_3=p-\langle p\rangle =p-p_0-x\xi (t)
-\frac{g\BX}{\gamma ^2+\alpha ^2}\,[\gamma +(\alpha \sin\alpha t-
\gamma \cos\alpha t)\,\re ^{-\gamma t}]
-\frac{g\BP}{\gamma ^2+\alpha ^2}\,[\alpha -(\gamma \sin\alpha t+
\alpha \cos\alpha t)\,\re ^{-\gamma t}].\nonumber\\ &&
\label{kamchadal}
\EEA
Here we have used Eqs.~(\ref{tulukan1}, \ref{tulukan2}, \ref{hu}), and
the average $\langle ...\rangle$ is taken by the noise operators 
$b$, $b^{\dagger}$ directly and through $\eta _0$. 
The quantity $\xi (t)$ was defined in Eq.~(\ref{sarancha}).
Notice that since we are interested here in the common state of the 
particle and the apparatus, the term connected with $\eta _1$
which gives rise to the last two terms 
in Eq.~(\ref{kamchadal}) appears as deterministic.
The gaussian quantum noise enters through
$B$, which is a $3\times 3$ symmetric matrix 
with the following elements:
\BEA
\label{hubilai1}
&&[B^{-1}]_{11}= \langle (X- \langle X\rangle )^2\rangle 
=[B^{-1}]_{22}=\langle (P- \langle P\rangle )^2\rangle
=\frac{\hbar}{2}~[1-\re^{-2\gamma t}]~{\rm coth}\frac{\hbar \alpha }{2T}, 
\\
\label{hubilai2}
&&[B^{-1}]_{12} = \langle X- \langle X\rangle ;P- \langle P\rangle\rangle =0, \\
\label{hubilai3}
&&[B^{-1}]_{13} = \langle X- \langle X\rangle ; p- \langle p\rangle \rangle =
\left\langle X- \langle X\rangle ; \int_0^t\d s~\eta _0(s) \right\rangle =
\frac{g\hbar}{2}\left (
\frac{\gamma -2\gamma \re ^{-\gamma t}\cos \alpha t+\gamma\re ^{-2\gamma t}}
{\gamma ^2+\alpha ^2}\right )~{\rm coth}\frac{\hbar \alpha }{2T},
\\
\label{hubilai4}
&&[B^{-1}]_{23} = \langle P- \langle P\rangle ;p- \langle p\rangle\rangle 
=\left\langle P- \langle P\rangle ; \int_0^t\d s~\eta _0(s) \right\rangle =
-\frac{g\hbar}{2}\left (
\frac{\alpha -2\gamma \re ^{-\gamma t}\sin \alpha t-\alpha\re ^{-2\gamma t}}
{\gamma ^2+\alpha ^2}\right )~{\rm coth}\frac{\hbar \alpha }{2T},
\EEA
\BEA 
[B^{-1}]_{33} = \langle (p- \langle p\rangle )^2\rangle &&=
\int _0^t\int _0^t\d s_1\d s_2 K_0(s_1,s_2)\nonumber\\
&&=\frac{g^2\hbar}{2}~{\rm coth}\frac{\hbar\alpha}{2T}~\left (
\frac{2\gamma t+1-\re ^{-2\gamma t}}
{\gamma ^2 +\alpha ^2}+
\frac{4\gamma (\gamma \re^{-\gamma t}\cos \alpha t-\alpha\re^{-\gamma t}\sin\alpha t
-\gamma )}
{(\gamma ^2 +\alpha ^2)^2}\right ).
\label{hubilai5}
\EEA

If we adopt the following notations
\BEA
&&\lambda (t) = [B^{-1}]_{11},\qquad
\zeta (t) =\frac{[B^{-1}]_{13}}{\lambda (t)}, \qquad
   \sigma (t) =\frac{[B^{-1}]_{23}}{\lambda (t)},\\
&& \Delta (t) =\frac{[B^{-1}]_{33}}{\lambda (t)}  -\sigma (t)^2-\zeta (t)^2,
\label{krim2}
\EEA
then we can write the matrix $B^{-1}$ as
\BEA
\label{13a}
B^{-1}=\la\left (\begin{array}{rrrr}
1\qquad & 0\qquad & \zeta & \\
0\qquad & 1\qquad & \sigma & \\
\zeta\qquad  & \sigma\qquad  & \Delta +\zeta ^2+\sigma ^2 &  
\end{array}\right ),
\EEA
and hence the matrix $B$ will read
\BEA
\label{13b}
B=\frac{1}{\la\Delta}\left (\begin{array}{rrrr}
\Delta +\zeta ^2\qquad & \zeta\sigma \qquad & -\zeta & \\
\zeta\sigma \qquad & \Delta +\sigma ^2 \qquad & -\sigma & \\
-\zeta  \qquad & -\sigma  \qquad & 1   &  
\end{array}\right )=
\frac{1}{\la}\left (\begin{array}{rrrr}
1 \qquad & 0\qquad & 0 & \\
0\qquad & 1\qquad & 0 & \\
0\qquad & 0\qquad & 0 &  
\end{array}\right )\, +\,
\frac{1}{\la\Delta}\left (\begin{array}{r}
-\zeta \\
-\sigma \\
1  
\end{array}\right )
\otimes\left (\begin{array}{rrr}
-\zeta &
-\sigma &
1  
\end{array}\right ).
\EEA
Eq.~(\ref{kr}) for $\Phi $ can
altogether be written in the explicit form:
\BEA
&&\Phi (X,P,x,p ;t~|~ X_0,P_0,x,p_0;0)=
\frac{1}{\sqrt{(2\pi)^{3}\lambda ^3\Delta }}
~\exp \left (
-\frac{1}{2\lambda } (L_1^2+L_2^2)
-\frac{1}{2\lambda \Delta } (L_3-\zeta  L_1-\sigma L_2)^2
\right ), 
\label{krim1}
\EEA
where $\la$, $\zeta$, $\sigma$, $\Delta$, $L_1$, $L_2$ and $L_3$ are defined by
Eqs.~(\ref{kamchadal0}~-~\ref{krim2}) as functions of time.

\subsection{Measurement of a variable with a discrete spectrum}
\label{disco}

The results reported so far were obtained for the measurement of the coordinate
$x$ which has a continuous spectrum. It is of clear interest to indicate
how the obtained results can be generalized for other specific situations,
e.g., for measurement of spin. Here we provide a simple remark, 
which will set our results in a more general context.

Let us assume that the interaction between the tested system and the
apparatus is still given by Eq.~(\ref{kamaz}) but now the tested system
is completely arbitrary, and $x$ in this equation refers to one of its 
observables. In particular, it can have a discrete spectrum. For simplicity 
we still neglect the self-Hamiltonian of the tested system.
In this general case, the complete Wigner function for the system
and the lowest level is no longer defined. Nevertheless, the intermediate 
Wigner function of Eq.~(\ref{kamaz1}) is still perfectly
defined. Recall that this function employs $(X,P)$ variables for the
lowest mode, but uses the matrix elements $(x',x'')$ in the eigenrepresentation
of the measured quantity $x$. As we show in Appendix A the intermediate
Wigner function corresponding to Eq.~(\ref{mr}) adequately
describes the general situation that we consider. Though the
complete density matrix of the measured
system and the lowest mode might also be used, the intermediate Wigner function
${\cal V}$ is a more convenient object to deal with.

An illustrative example for the measurement of an observable with a discrete
spectrum is the spin-boson Hamiltonian \cite{weiss}:
\BEA
H_{\rm I}=\frac{1}{2}g\sigma _zX,
\EEA
where the measured observable $\frac{1}{2}\sigma _z$ 
is the $z$-component of spin for the tested system.

The situation of a discrete spectrum measurement will also be
encountered below for our original model in spite of the continuity of the 
coordinate $x.$ Actually, we shall consider an initial density operator $\rho (0)$
involving two distinct values of $x$ only, a situation which does not differ
much from a genuine discrete spectrum.

\renewcommand{\thesection}{\arabic{section}}
\section{Ideal measurement: Postmeasurement states}
\setcounter{equation}{0}\setcounter{figure}{0} 
\renewcommand{\thesection}{\arabic{section}.}

In this section we will be interested by the postmeasurement situation.
Let us first resume the {\it conditions of ideality} of the measurement that
we have encountered above. The relaxation time of the apparatus should 
be small compared to the duration of the measurement:
\BEA
\label{jihad0}
\gamma \th \gg 1.
\EEA
The coupling with the bath should however be small on the scale of 
the dynamical time (\ref{imp}):
\BEA
\label{k0}
\gamma \ll \alpha =\frac{|\mu |}{\hbar}.
\EEA
On the other hand, if we denote by $\tilde{x}$ a typical value of the 
coordinate to be measured, 
\BEA
\label{tilde}
\tilde{x}^2 = {\rm tr}(\rho (0)x^2),
\EEA
its coupling with the apparatus should produce
a finite condensate density, which according to Eq.~(2.20) is
expressed as
\BEA
\label{k1}
\frac{\hbar g^2\tilde{x}^2}{\mu ^2}={\cal O}(N).
\EEA
The fact that the bath ensures Bose condensation, but that the condensate 
density remains dominated by the coupling with the particle, imposes the 
condition (\ref{mazarishari}), that is 
\BEA
\label{gustavadolf}
1\gg \frac{|\mu |}{T}\gg \frac{1}{N}.
\EEA
To fix ideas we shall assume in the following that 
\BEA
\label{kabul0}
\frac{|\mu |}{T}={\cal O}\left (\frac{1}{\sqrt{N}}\right ),
\EEA
with $T$ finite in the thermodynamical limit. This will imply, 
from Eq.~(\ref{k1}), that the coupling constant $g$ is finite.
Finally, if $\tilde{p}$ is the characteristic value of the particle momentum,
the approximate conservation of $x$ during the measurement means that 
\BEA
\frac{\tilde{p}}{m}\,\th\ll \tilde{x}.
\label{jihad1}
\EEA
or $m\gg g^2\theta ^3$ as shown in Appendix C. We expect that under these conditions both
the tested system and the Bose gas have reached at time $\theta $ a 
quasistationary state that we wish to study.

By using the condition (\ref{jihad0}) we first recall
that the excited states become decoupled and thermalized with the bath, since
their Wigner function at time $\th $ deduced from Eq.~(\ref{plumbum0}) 
reads
\BEA
W_k(X_k,P_k)=\frac{2}{{\rm coth}\frac{\varepsilon _k+\hbar\alpha}{2T}}
\exp \left[
-\frac{1}{\hbar ~{\rm coth}\frac{\varepsilon _k+\hbar\alpha}{2T}}
(X_{k}^2+P_{k}^2)
\right ].
\EEA

The transition kernel $\Phi$ between times $0$ and $\th$ can be obtained from 
Eqs.~(\ref{kamchadal0}~-~\ref{krim2}, \ref{krim1}). Under
conditions (\ref{jihad0}, \ref{k0},
\ref{gustavadolf}, \ref{jihad1}) its various ingredients reduce to
\BEA
\label{kipchak2}
&&\lambda (\th ) =
\frac{\hbar}{2}~{\rm coth}\frac{\hbar \alpha }{2T}
\simeq \frac{T}{\alpha},\qquad
\zeta (\th )=0, \qquad \sigma (\th ) =-\frac{g}{\alpha},\\
&& \Delta  (\th )=\frac{2g^2\gamma\th}{\alpha ^2}, \\
&& L_1 (\th )= X-\frac{gx}{\alpha}, \qquad L_2(\th )=P, \qquad
L_3(\th )=p-\Bp -\frac{xg^2\th}{\alpha} -\frac{g}{\alpha}\BP ,\\
&& L_3-\zeta L_1-\sigma L_2=p-p_0+\frac{g}{\alpha}(P-P_0)-
\frac{xg^2\th}{\alpha},
\label{kipchak1}
\EEA
where we have used $\xi (\th )=g^2\th /\alpha$ as follows from 
Eq.~(\ref{sarancha}).
We note that Eqs.~(\ref{gustavadolf}, \ref{kabul0}) imply
\BEA
\label{timur0}
N\hbar \gg \la (\th ) \gg \hbar,\qquad \la =\hbar {\cal O}(\sqrt{N}),
\EEA
repectively,
and that Eqs.~(\ref{jihad0}, \ref{k1}) imply
\BEA
\label{timur1}
\Delta (\th )\gg \frac{N\hbar}{\tilde{x}^2},
\EEA
while $g\tilde{x}/\alpha $ is of order $\sqrt{N\hbar}$.

The postmeasurement state of the tested particle and the apparatus will be
investigated below in three steps. First we will discuss the
partial state of the apparatus and that of the particle. Later we turn to the
global state.

\subsection{Apparatus}

As seen, the apparatus itself behaves as a Bose gas, subject 
to a source field proportional to $gx(0)$.
Owing to Eqs.~(\ref{timur0}, \ref{timur1}) the variance $\la\Delta$ of $L_3$
in Eq.~(\ref{krim1}) is much larger than $(N\hbar/\tilde{x})^2$. 
We can therefore readily trace out the tested particle from Eq.~(\ref{mr}, \ref{krim1})
by integrating over both $p$ and $p_0$. The factor $w(x,p_0;0)$
of Eq.~(\ref{mr}) thus generates the probabiilty density:
\BEA
\label{kuzma}
\int \frac{\d p_0}{2\pi\hbar}~w(x,p_0;0)= \langle x|\rho (0)|x\rangle
\EEA
for the coordinate $x$ in the initial state of the tested particle. 
The resulting expression of the Wigner function of the apparatus at time 
$\th$ has the expected form
\BEA
\label{full}
W(X,P;\th )=\int \d x~\langle x|\rho (0)|x\rangle ~W_x(X,P),
\EEA
\BEA
W_x(X,P)=\frac{\hbar}{\lambda}
\exp \left [
-\frac{1}{2\lambda}
\left (X-\frac{gx}{\alpha}\right )^2
-\frac{1}{2\lambda}
P^2 \right ].
\label{wiko}
\EEA
The Wigner function $W_x(X,P)$, where $\la =T/\alpha ,$ describes 
the quantum Gibbs distribution 
of the apparatus at temperature $T$ and chemical potential 
$\mu =-\hbar\alpha$, with a classical source $J=\sqrt{\hbar/2}\,gx$.

For each possible value of the coordinate $x$ of the particle the apparatus
is thus in an equilibrium state, with an order parameter proportional to
$x$. In spite of the quantum nature of the variable $x(0)$ which is
governed by the initial density operator $\rho (0)$, it acts on the apparatus as
a {\it classical random} object. 
We can understand this classical feature by noting that because $x(0)=x(t)$
during the process, $a(t)$ and $a^{\dagger}(t)$ commute with it:
$[a(t),x(0)]=0$, $[a^{\dagger}(t),x(0)]=0$.
Therefore, the situation is very similar to that considered in 
section \ref{bose}, with $J=gx(0)\sqrt{\hbar/2}$.
However, there is a subtle point, since the field $J$ is now random. Its
quantum randomness arises from the initial state of the tested particle, which
in general is not an eigenstate of the coordinate operator.
Since the off-diagonal part of $\rho (0)$ disappears owing to the large 
size of $\la (\th )\Delta (\th )$, Eq.~(\ref{full}) shows that
the quantum nature of the randomness is suppressed.
It is seen as well that the resulting classical 
randomness is quenched, which means that 
all extensive quantities have to be calculated for a fixed field 
and then averaged at the last step.

Notice that for the continuous spectrum a certain difficulty may arise if
(for example) the initial state of the particle is an eigenstate of the
coordinate: $\langle x'|\rho (0)|x''\rangle =\delta (x'-x_0)\delta (x''-x_0)$,
in which case the quantity (\ref{kuzma}) diverges.
There are several standard ways to overcome this difficulty
\cite{landau}. The simplest one is to consider a gaussian packet
$\langle x|\psi\rangle =
(2\pi\epsilon )^{-1/4}\exp[-(x-x_0)^2/(4\epsilon)]$
instead of a precise eigenstate of the coordinate.
This state has a normalizable Wigner function,
\BEA
\label{foma}
w(x,p;0)=2\exp \left [
-\frac{(x-x_0)^2}{2\epsilon}-\frac{2\epsilon ~p^2}{\hbar ^2}
\right ],
\EEA
and we can let $\epsilon\to 0$ in the
last stage of calculations. Further on we will always assume 
that this procedure is implied when necessary.
This justifies the replacement of (\ref{kuzma}) by $\delta (x-x_0)$.

Since the apparatus is a non-ergodic system, Eq.~(\ref{full})
means that it will occupy with a priori
probabilities $\langle x|\rho (0)|x\rangle $ a
state which is determined by the initial value of the coordinate.
If the initial state of the particle is an eigenstate
of the coordinate operator the field is not random, and
there is only one state to which the apparatus can relax. So  
in this case we have a definite prediction, as it should be.

In each state (\ref{wiko}) of the apparatus, the momentum $P$
associated with the lowest mode fluctuates around the value $0$
exactly as in the gibbsian state imposed by the bath, but its coordinate
$X$ is shifted by $xg\hbar /|\mu |$. This shift is of order $\sqrt{N\hbar}$
according to Eq.~(\ref{k1}), and it produces a finite shift in the density
\BEA
\label{fatah}
\frac{N_{\rm c}}{V}=\frac{1}{2\hbar V}~\langle X^2+P^2-\hbar\rangle\simeq
\frac{\hbar g^2x^2}{2\mu ^2V}
\EEA
of the condensate. The statistical fluctuation of $X$, equal to $\sqrt{\la}$,
is small compared to $\langle X\rangle \propto \sqrt{N\hbar}$ owing to the 
first condition in Eq.~(\ref{timur0}), namely $\la\ll N\hbar$.

The second condition in Eq.~(\ref{timur0}), namely $\la\gg\hbar$, entails
that the variables $X$ and $P$ behaves as {\it classical} random variables, 
which is a natural requirement for the pointer variable of an apparatus.
Compared to the shift $\left\langle X\right\rangle$ their
fluctuations are of relative order $N^{-1/4}$ if we choose
$\lambda =\hbar {\cal O} (\sqrt N).$

\subsubsection{Amplification and registration}

We have just seen that the interaction of the tested particle and the
apparatus results in a {\it macroscopic} change in the condensate density: 
It fixes the expectation value $\langle X\rangle$ to $xg\hbar/|\mu |$
within fluctuations which are small in relative value. This large effect is a
consequence of the condition (\ref{k1}). The coupling $g$ is sufficiently 
large to produce a shift in the number $N_{\rm c}$, which is of the same order 
as the total particle number. However, it is sufficiently small so that
the contribution of the interaction Hamiltonian $H_{\rm I}$ to the energy of
the apparatus is negligible.

The amplification of the influence of the tested system on the apparatus is due
here to the smallness of $|\mu |$. The bath, which imposes on the apparatus the
condition (\ref{gustavadolf}), prepares it before the measurement in a
state where the condensate density is not yet finite, but where the
smallness of $|\mu |$ makes the apparatus very sensitive to a source
coupled to $X$. By making successive macroscopic observations of the
value of $X$, one can
then find the statistics of $x(0)$ through $\langle x|\rho (0)|x
\rangle$. There is one-to-one correspondence between $\langle X\rangle$
and $x(0)$, without bias because the various values of the order
parameter $\langle X\rangle$ yield identical values for the energy as well as
for the entropy.

A peculiarity of the model comes from the fact that the amplification factor
$g/|\mu |$ depends on the coupling constant $g$ and on the chemical potential 
of the bath. These quantities, which are both small, need to be known to let
us determine $x(0)$ through $\langle X\rangle$. 

Moreover, if we wish to {\it register} the result of a measurement, which is
the value reached by $X$ at the time $\th$ when the interaction $g$ is
switched off, we need to imagine that the exchange of bosons between the
bath and the apparatus is also switched off at the same time $\th$.
The overall density of bosons as well the condensate density thereafter
remain fixed in the apparatus, which is in canonical equilibrium after
the time $\th$. 
Another theoretical procedure to freeze the condensate density at the value (\ref{fatah})
would consist in switching off the coupling $\gamma$ before the end
of the measurement. The fate of the tested particle after the
time $\theta $ is considered in Appendix B.

\subsubsection{Robustness}

We have explained how our apparatus realizes amplification
of weak signals. This is only half of the way towards a good 
information storing device, because we yet should see whether another 
important property which is {\it robustness} is satisfied. 
In other words, if under influence
of a weak field the apparatus has relaxed to a definite state, then 
what is the probability that it will leave this state spontaneously?
If this transition probability is small, and can be made as small
as it is desired, then the property of robustness is present.

Let us assume that the apparatus has been brough into a state
with
\BEA
\langle X\rangle =\frac{gx}{\alpha}. 
\EEA
In this state the apparatus has Wigner function $W_x $ and density matrix $R$.
We wish to calculate the transition probability to
another state $R'$ associated with $x'$
under the effect of some perturbation.
If these states were pure, the transition probability would read as usual:
\BEA
{\rm Pr}(x\to x ')=
{\rm tr}(RR ').
\EEA
For mixed states we use the same formula in terms of the overlap:
\BEA
{\rm Pr}(x\to x')\propto {\rm tr}(RR ')=
\int \frac{\d X~\d P}{2\pi\hbar}~W_x(X,P)W_{x'}(X,P).
\EEA
To be normalized this expression should be divided by ${\rm Pr}(x\to x)$. 
Using Eq.~(\ref{wiko}) one gets
\BEA
{\rm Pr}(x\to x ')=
\exp \left[
-\frac{g^2(x-x' )^2}{4\la\alpha ^2 }
\right ].
\label{tanjer}
\EEA
It is clear that above the phase transition point,
when $\la\alpha ^2$ is finite, this transition
probability is of order one, so that no robustness is present as 
was to be expected.

Let us consider the situation below phase transition
where $\la\sim T/\alpha$. We then have
\BEA
{\rm Pr}(x\to x ')=
\exp \left[
-\frac{\hbar g^2(x-x' )^2}{4T |\mu |}
\right ].
\label{ichi}
\EEA
According to Eq.~(\ref{gustavadolf}) 
the exponent is of order $-N|\mu |/T$. With Eq.~(\ref{kabul0})
this exponent behaves as $-\sqrt{N}$ in the thermodynamical limit, 
provided $x$ and $x'$ differ by a quantity which remains finite as 
$N\to \infty$. The probability therefore vanishes, as it should.

The fact that the overlap between states of the apparatus associated 
with different values of $x$ is negligible also expresses that different
positions of the pointer variable constitute {\it exclusive} events.

\subsubsection{Accuracy of measurement}

The robustness reflects stability of the apparatus with respect to 
external perturbations. Another quantity, the {\it accuracy},
characterizes the strength of the noise due to the initial 
uncertainty of the tested particle, and due to spontaneous thermal 
fluctuations induced by the bath.
One can estimate the accuracy of the measurement by evaluating the
following quantity
\BEA
\Sigma =\frac{\{\langle X^2 \rangle\}_{\rm av} 
-\{\langle X\rangle ^2\}_{\rm av} }
{\{\langle X\rangle ^2\}_{\rm av} },
\EEA
where the average $\langle ...\rangle$ is taken with respect to the 
state of the apparatus, and where $\{ ... \}_{\rm av}$ denotes the 
average over the initial distribution of the particle.
This is the signal-to-noise ratio, and $\Sigma\ll 1$ corresponds to a
good measurement of $X$.
Having used Eqs.~(\ref{urumchi1}, \ref{urumchi2},
\ref{gustavadolf}), one finally gets
\BEA
\Sigma =\frac{T|\mu |}{\hbar g^2\{ x^2\}_{\rm av}}.
\EEA
In the region where the condensational phase transition exists
this quantity is small as $T/(|\mu |N)={\cal O}(N^{-1/2})$, 
provided that the thermodynamical 
limit is taken 
and that $\{ x^2\}_{\rm av}$ is finite. The accuracy is thus
governed by thermal noise.

Note also that apart from the above uncertainty the derivation
of the measured quantity $x$ from the pointer variable $X$ by
means of Eq.~(4.19) involves the ratio $g/\mu .$ The accuracy
of the measurement is, of course, spoiled if the coupling constant $g$
and the chemical potential of the bath are not controlled with
precision.

\subsection{Tested particle}
\label{harun}

Let us now consider the partial state of the tested particle.
At time $\th$ we can find this state by tracing out the apparatus 
from Eq.~(\ref{mr}) using the approximations (\ref{kipchak2}~-~\ref{kipchak1}).
In this calculation we first note that the memory about the initial value 
$X_0$ is lost. The variable $P_0$ enters through the last term of
the exponent of $\Phi$, which at the time $\th$ reads:
\BEA
-\frac{1}{2\la\Delta}(L_3-\zeta L_1-\sigma L_2)^2
=-\frac{1}{2\la\Delta}\left(p-p_0+\frac{g}{\alpha}(P-P_0)-
\frac{xg^2\th}{\alpha}\right)^2.
\label{kali}
\EEA
Since the apparatus is nearly in equilibrium at both times $0$ and $\th$,
$P$ and $P_0$ are of order $\sqrt{\lambda}=\sqrt{T/\alpha}$. We can thus neglect
the term depending on the apparatus in the bracket of Eq.~(\ref{kali}),
because
\BEA
\frac{1}{2\lambda\Delta}
\frac{g^2}{\alpha^2}
(P-P_{0})^2=
\frac{\alpha^3}{4g^2T\gamma\th}
\frac{g^2}{\alpha^2}
\frac{T}{\alpha}=\frac{1}{4\gamma\th}
\EEA
is small. This means that the overall system forgets about the initial state
of the apparatus for $\gamma\th\gg 1$. We can therefore readily integrate over
the initial state $W(X_0,P_0;0)$ of the apparatus, then over the variables
$X$ and $P,$ which yields
\BEA
\label{imam}
w(x,p;\theta ) = \int \d \Bp ~w(x,\Bp ;0)
\frac{1}{\sqrt{2\pi \la\Delta}}
\exp \left [
-\frac{1}{2 \la\Delta}\left (p-\Bp -x
\frac{g^2\th}{\alpha} \right)^2
\right ].
\EEA
Due to the large value of $\la\Delta$ the exponential factor in 
Eq.~(\ref{imam}) is nearly constant. Indeed, we have
\BEA
\label{misra}
\frac{\la\Delta\tilde{x}^2}{\hbar ^2}=
\frac{2\hbar g^2\tilde{x}^2}{\mu ^2}
\frac{T}{|\mu |}
~\gamma \th=\gamma \th~{\cal O}(N^{3/2}),
\EEA
where $\tilde{x}$ is defined by Eq.~(\ref{tilde}),
so that the only effect of this exponential is to produce a cut-off which 
ensures the normalization of $w(x,p;\th )$. Otherwise,  $w(x,p;\th )$ is
practically independent of $p$, and its dependence on $x$ is the same as
that of the probability density $\langle x|\rho (0)|x\rangle$ as expected.

The density matrix in the $x$-basis associated with the Wigner function
(\ref{imam}) is given by 
\BEA
\langle x'|\rho (\theta ) | x'' \rangle =
\langle x'|\rho (0) | x'' \rangle
\exp \left [
-\frac{\la\Delta}{2\hbar ^2}(x'-x'')^2+~\frac{{\rm i}g^2\th}{2|\mu |}(x'^2-x''^2)
\right ].
\label{shuravi}
\EEA
The large value of $\la\Delta$ ensures that it practically reduces 
to the diagonal part of $\rho (0)$ in the basis $x$.

\subsubsection{Decoherence time}
\label{deco}
The above expressions for $w(\th )$ or $\rho (\th )$ show that the 
decoherence of the state of the tested particle with respect to $x$
has been achieved at the time $\th$.

In order to understand how and when 
this {\it decoherence} takes place during the
interaction process between times $0$ and $\th$, we return to the equations
of motion of the tested particle which are given by 
Eqs.~(\ref{kismet}, \ref{vavilon77}, \ref{hu}).
Both terms $\eta _0$ and $\eta _1$ in Eq.~(\ref{hu})
should here be treated as noise, since we eliminate the apparatus. 
The fluctuations of this noise are given by Eqs.~(\ref{kalgata1}, 
\ref{kalgata2}). 
Altogether we get for the Wigner function $w(x,p;t)$ 
of the tested particle:
\BEA
w(x,p;t) = \int \d \Bp ~w(x,\Bp ;0)
\frac{1}{\sqrt{2\pi d(t)}}
\exp \left [
-\frac{1}{2d(t)}(p-\Bp -x\xi (t))^2
\right ],
\label{molla}
\EEA
where we define $d(t)$ by
\BEA
d(t)=
\int _0^t\int _0^t\d t_1~\d t_2~K(t_1,t_2)=
\int _0^t\d t_1~\d t_2~K_0(t_1,t_2)+\frac{g^2\hbar}{2}
{\rm coth}\frac{\hbar\alpha}{2T}\left (
\frac{1-2e^{-\gamma t}\cos \alpha t+e^{-2\gamma t}}{\gamma ^2 +\alpha ^2}
\right ),
\EEA
and $\int _0^t\int _0^t\d t_1~\d t_2~K_0(t_1,t_2)$ 
is given by Eq.~(\ref{hubilai5}).
By using Eqs.~(\ref{hubilai3}, \ref{hubilai5}) we find
\BEA
d(t) = [B^{-1}]_{33}+ \frac{g}{\gamma}[B^{-1}]_{13}.
\EEA

Having translated to density matrix with help of formulas
(\ref{rhoW}, \ref{Wrho}) one gets the evolution as
\BEA
\langle x'|\rho (t) | x'' \rangle =
\langle x'|\rho (0) | x'' \rangle
\exp \left [
-\frac{d(t)}{2\hbar ^2}(x'-x'')^2+\frac{{\rm i}\xi (t)}{2\hbar}(x'^2-x''^2)
\right ].
\label{khorezm}
\EEA
It is seen that the diagonal elements of the density matrix
(those for which $x'=x''$) are not changed at all, whereas the
off-diagonal ones are damped with rate $d(t)/\hbar ^2$.
In other words, the density matrix of the tested particle
tends to the mixture formed by eigenstates of the coordinate operator. 
Indeed, the decoherence factor $d(t)$, which measures how the density matrix
is squeezed in terms of $x'-x''$, increases from $0$ to $\infty$ as the time
$t$ goes from $0$ to $\infty$. Its asymptotic forms 
at short and long times are derived from Eqs.~(\ref{hubilai3}, \ref{hubilai5})
as
\BEA
\label{mustafa1}
&&d(t)= \frac{ g^2T}{\alpha }~t^2, 
\quad 
\gamma t \ll \alpha t\ll 1,
\\
&&d(\th )=\frac{2 g^2T}{\alpha ^3}~\gamma\th, 
\quad 
\gamma\th\gg 1. 
\label{mustafa2}
\EEA
For long times $\gamma\th\gg 1$ the
exponent of Eq.~(\ref{khorezm}) behaves as
\BEA
\label{mustafa3}
&&
\frac{d(\th )}{2\hbar ^2}(x'-x'')^2
\sim \frac{\hbar g^2(x'-x'')^2}{\mu ^2}~\frac{T}{|\mu |}~\gamma\th
=\gamma\th ~{\cal O}(N^{3/2}),
\EEA
and we recover as in Eq. (\ref{shuravi}) the 
strong damping of the off-diagonal terms,
which allows us to obtain of the proper state of the tested particle after
the measurement. 

An analogous investigation for short times $\gamma t\ll \alpha t\ll 1$
can be carried out using Eq.~(\ref{mustafa1}):
\BEA
\label{alibaba1}
\frac{d(t )}{2\hbar ^2}(x'-x'')^2
\sim \frac{\hbar g^2(x'-x'')^2}{\mu ^2}~\frac{|\mu |}{T}~
\left(\frac{Tt}{\hbar}\right)^2=
\left(\frac{Tt}{\hbar}\right)^2~{\cal O}(N^{1/2}).
\EEA
This quantity can be much larger than $1$ even in the region of short times
$t$ for which $\alpha t=|\mu |t/\hbar\ll 1$, provided
\BEA
N^{-1/4}\ll\frac{Tt}{\hbar}\ll \frac{T}{|\mu |}={\cal O}(N^{1/2}).
\EEA
The off-diagonal terms of $\rho (t)$ thus disappear at the very
beginning of the interaction of the particle with the apparatus, 
after a delay of order $N^{-1/4}$. The characteristic 
time
\BEA
\tau \sim \frac{\hbar}{TN^{1/4}}
\label{gallo}
\EEA
over which the reduction of the state of the tested particle takes place
is thus {\it much shorter} than the {\it duration} $\th$ of the measurement, since
\BEA
\th\gg \frac{1}{\gamma}\gg\frac{\hbar}{|\mu |}\gg \tau .
\label{turkestan}
\EEA
The first inequality in Eq.~(\ref{turkestan}) ensures the
relaxation of the apparatus. The second inequality ensures that the
stationary state will be gibbsian. The third inequality indicates that
the decoherence of the tested particle 
takes place on a time-scale much smaller that the dynamical
time $\hbar/|\mu |$ given by Eq.~(\ref{imp}).
Note also that $\hbar/T$ itself is an important characteristic time-scale
in quantum statistical physics, which characterizes the relevance of
quantum versus thermal effects. As shown in Eq. (\ref{gallo}) the
macroscopic size of the apparatus reduces these quantum effects
by a factor $N^{1/4}.$

The {\it reduction time} $\tau$, which in textbook
discussions is either taken to be zero or identified with the 
{\it measurement time} $\th $ itself, is definitely different from the latter
in our model. Another interesting  
aspect of this difference is that, at those short time-scales given by 
Eq.~(\ref{gallo}), the change in the apparatus is still negligible.
Indeed, the variables of the bath have not yet changed at those scales, 
since the characteristic time over which the equilibrium between 
the bath and the apparatus sets in is $1/\gamma$.
On the other hand, the energy associated with the particle has
as well not changed, because the influence of the kinetic
energy can be neglected, and the coordinate $x(t)$ remains constant.
In fact we shall study in the next subsection the change in
momentum associated with the second, imaginary term in the
exponent of Eq.~(\ref{khorezm}). According to Eqs.
(\ref{mustafa1}) and (\ref{grek}) below, this term is negligible
compared to the first one for $\alpha t\ll 1$ since their ratio
for such short times is of order $\hbar \xi (t)/d(t)\sim \alpha t
|\mu |/6T=\alpha t~{\cal O} (N^{-1/2}).$ Altogether, this means that
Eqs.~(\ref{alibaba1}, \ref{gallo}, \ref{turkestan}) 
provide an example of a situation where the reduction of the state of
the tested particle occurs long before achievement of the
measurement process, and without any energy cost, although this
reduction is a consequence of the interaction with the apparatus.

\subsubsection{Backreaction of the apparatus}
\label{tri}

The post-measurement Wigner function (\ref{imam}) of the tested particle
involves a shift $xg^2\th /\alpha$ of the momentum, that we have not yet 
discussed. The density matrix (\ref{shuravi}) accordingly exhibits 
oscillations within the small region in $x'-x''$ where 
$\langle x'|\rho (t) | x'' \rangle$ is significant. Let us evaluate the
order of magnitude of the corresponding average momentum:
\BEA
\langle p\rangle = \frac{\hbar x g^2\th }{\mu}=
\frac{\hbar g^2\tilde{x}^2\th }{\mu ^2}~\frac{\alpha}{\gamma}~
\frac{x\hbar}{\tilde{x}^2}~\gamma \th 
=\frac{\hbar}{\tilde{x}}~\frac{\alpha}{\gamma}~\gamma \th
~{\cal O}(N)\gg \frac{\hbar}{\tilde{x}}~{\cal O}(N).
\EEA
The fluctuations of $p$, of order $\sqrt{\la\Delta}=\sqrt{d(\theta)},$
are also large,
because the final state is localized in the $x$-space. They are
expressed
by Eq.~(\ref{misra}). Although large, the value of $\langle p\rangle$
would be ineffective if it were smaller than
these fluctuations. However, the ratio of the shift to the fluctuations
is given at the time $\th$ by
\BEA
\label{muk}
\frac{\langle p\rangle ^2}{\langle p^2\rangle -\langle p\rangle ^2}
=\frac{\hbar g^2x^2}{2\mu ^2}~\frac{T}{|\mu |}~
\left (\frac{\alpha}{\gamma}\right )^2~\gamma \th \gg
\left (\frac{\alpha}{\gamma}\right )^2~\gamma \th ~{\cal O}(N^{1/2}),
\EEA
a large number in the thermodynamical limit. 

The shift of $\langle p\rangle$ is therefore an important effect 
of the measurement. Let us look how this shift increases at short times.
By using Eqs.~(\ref{molla}, \ref{sarancha}) we find:
\BEA
\label{grek}
x\xi (t )\sim \frac{g^2\alpha x t^3}{6}, \qquad \gamma t\ll \alpha t\ll 1.
\EEA
At the time $\tau$ when decoherence is being achieved, the shift
(\ref{grek}) is of order
\BEA
x\xi (\tau )\sim \frac{\hbar g^2x^2}{\mu ^2} \left (
\frac{\mu}{T}
\right )^3~N^{-3/4}~\frac{\hbar}{x} =\frac{\hbar}{x}~{\cal O} (N^{-5/4}).
\EEA
We see that the change in $\langle p\rangle$ begins to be significant
long after decoherence has taken place. This is consistent with the fact
anticipated at the end of section \ref{deco}, that very little is yet
changed in the apparatus at the time $\tau$.

The ratio (\ref{muk}) is of order $1$ at a time $\tau _1$ such that 
$[x\xi (\tau _1)]^2=d(\tau _1)$, where the shift becomes
comparable with the fluctuation.
Using Eqs.~(\ref{mustafa1}, \ref{grek}) one finds
\BEA
\tau _1 = \frac{\hbar}{T}N^{3/8}, \qquad \tau\ll\tau _1\ll \frac{1}{\alpha}.
\EEA
Here as in $\tau$, the time-scale is given by $\hbar /T$, but now the
thermodynamical limit produces an enhancement.

The large shift of $\langle p\rangle$ can be attributed to the
interaction process which takes place between the apparatus and the
particle in the time interval $\tau _1$, $\th$.
During this period, the particle acts upon the lowest mode so as to
drive it
towards a state with a finite density of the condensate. In response
this large effect produces a boost in the average momentum
$\langle p\rangle$.
The phenomenon can be traced back to the second term in the r.h.s. of 
Eq.~(\ref{vavilon7}) for $p$.  The factor $\chi (t)$ which enters this term
describes a deterministic effect produced on the particle by the 
apparatus in contact with the bath. The increase of
$|\langle p\rangle |$ is thus a cumulative
effect of friction, which accompanies the rise of $|\langle X\rangle |$.

Altogether the interaction of the tested particle with the apparatus
produces on this particle two effects. It first reduces the state,
suppressing the 
off-diagonal terms in $x$ during the time $\tau $. Later on, between
the times $\tau _1$ and $\th ,$ it yields a large value to the average 
momentum without spreading the distribution in $x$, as seen 
in Eqs.~(\ref{imam}, \ref{molla}). This second effect is probably
connected with specific features of our model, namely, the
choice of the apparatus and of its order parameter, and the form of the
interaction Hamiltonian between the apparatus and the tested particle.
It is also related to the existence
of the continuous spectrum for $x$, which allows the rapid oscillations
exhibited by Eqs.~(\ref{shuravi}, \ref{khorezm}). 

One may wonder whether the large value of $\langle p\rangle$
reached at the time $\theta $ is compatible with our hypothesis
that the Heisenberg operator $x(t)$ has remained practically
constant over the time interval $0,\theta .$ We show in Appendix
C that, contrary to $p(t),$ the equation of motion for $x(t)$
contains no systematic drift term arising from the coupling with
the apparatus and hence with the bath; it involves only a noise
term which does not affect much $x(t).$ We were thus entitled to
neglect the variations of $x$ between the times $0$ and $\th .$

\subsubsection{Einstein-Podolsky-Rosen experiment and speed of quantum
signals}
The above analysis allows us to discuss an experiment of the
Einstein-Podolsky-Rosen (EPR) type. Let us suppose that the tested
system consists of two particles denoted by $1$ and $2$. They do not
interact for $t>0$, but they did interact in the past, which is reflected
in an entangled wave-function of the tested system at the initial time $t=0$:
\BEA
\label{deon}
\rho (0)=|\psi\rangle\langle \psi |,\qquad
|\psi\rangle = \sum _k\alpha _k|x_k\rangle |y_k\rangle ,
\EEA
where $|x_k\rangle$ are eigenfunctions of the operator $x$ 
for the first particle, and $|y_k\rangle $ are arbitrary normalized,
not necessarily orthogonal
functions in the Hilbert space of the second particle. 
As indicated by Eq.~(\ref{foma}) a small dispersion should be allowed
for $x$ so as to normalize $|x_k\rangle$.
The measurement of
the observable $x$ is realized as above, namely the first 
particle couples through
its operator $x$ with the apparatus as expressed by Eq.~(\ref{kamaz}).
However, the second particle {\it does not} interact with the apparatus.
Eqs.~(\ref{khorezm}, \ref{alibaba1}) take place
as above with the slight difference that 
$\langle x'|\rho (t) | x'' \rangle$, 
$\langle x'|\rho (0) | x'' \rangle$ are matrices in the Hilbert space of the 
second particle. In particular, the reduction (collapse) of the initial state
occurs on the time-scale predicted by
Eqs.~(\ref{alibaba1}, \ref{gallo}), and it now provides:
\BEA
\rho (\tau )\simeq \sum _k|\alpha _k|^2\, |x_k\rangle |y_k\rangle \, \langle y_k|
\langle x_k|.
\EEA
The remarkable feature of quantum mechanics is that although only one
subsystem is involved in the measurement, the total state of the tested 
system is reduced. 

This general analysis can be illustrated by the standard example of two
spins produced in the singlet state by the decay of an object with
angular momentum $0$. The initial density operator (\ref{deon}) is
\BEA
\langle s_1s_2|\rho (0) | s_1's_2' \rangle =\frac{1}{2}\delta _{s_1+s_2,0}~(\,
\delta _{s_1,s_1'}\delta _{s_2,s_2'}-
\delta _{s_1,s_2'}\delta _{s_2,s_1'}).
\EEA
As above we perform measurement on the $z$-component of the first spin
only. For large $N$ Eq.~(\ref{alibaba1}) leads to reduction $s_1=s_1'$
after a delay $\tau$. This automatically implies $s_2=s_2'$:
\BEA
\langle s_1s_2|\rho (\tau ) | s_1's_2' \rangle \simeq \frac{1}{2}
~\delta _{s_1+s_2,0}~\delta _{s_1,s_1'}~\delta _{s_2,s_2'}.
\EEA

Let us give some quantitative estimate of the characteristic reduction
time-scale $\tau$. For a temperature $T\simeq 1~{\rm K}$ we obtain
\BEA
\tau =10^{-11} N^{-1/4} \,{\rm s}.
\EEA
Estimating $N\sim 10^{24}$ for a macroscopic system,
we get $\tau \sim 10^{-17}\,{\rm s}$. For a distance of $1\,{\rm m}$
between spins
this would lead to a speed of order $10^{17}\,{\rm m/s}$. Of course,
this does not mean that there is an information transfer at this
speed, but only a change in our knowledge through $\rho $ of the
quantum correlations of the two spins. Indeed, as we noticed at the end
of section 4.2a the energy of the system is constant for times of
order $\tau .$ 

\subsection{The common state of the tested particle and apparatus}

In the evaluation of the
common Wigner function (\ref{mr}) of the particle and the apparatus,
we use for the transition kernel
$\Phi$ the approximate expressions (\ref{kipchak2}~-~\ref{kipchak1})
as above. We recall that in
the limit $\gamma \th \gg 1$ the variance of $p$,

\BEA
d(\th )=\la (\th )\Delta (\th )=\frac{2g^2T}{\alpha ^3}~\gamma\th =
\frac{\hbar ^2}{\tilde{x}^2}~\gamma\th ~{\cal O}(N^{3/2}),
\EEA
is large as well as $\la (\th )=T/\alpha$. We saw at the
beginning of Section 4.2 that this implies a loss of memory about
the initial state $W(X_0,P_0;0)$ of the apparatus. In Eq.~(\ref{mr}) we
thus integrate over the initial Wigner function (\ref{kaban}) and find
\BEA
{\cal W}(X,P,x,p;\th )=\int \d \Bp ~\Psi (X,P,x,p;\theta |x,p_0)w(x,p_0;0),
\label{pr1}
\EEA
where
\BEA
\label{werner1}
\Psi (X,P,x,p;\th ~|~x,p_0)&&=
\frac{\hbar }{\lambda (\th ) \sqrt{2\pi d(\th )}}\times \nonumber\\
&&\exp\left [
-\frac{1}{2\lambda }\left (X-x~\frac{g}{\alpha }
\right )^2-\frac{1}{2\lambda }P^2
-\frac{1}{2d(\th )}
\left (p-\Bp -\xi (\th )x +\frac{g}{\alpha}P\right )^2
\right ].
\EEA
For $d(\th )\rightarrow \infty $ the expression (\ref{pr1})
reduces to the product of the Wigner function $W_x(X,P)$ of the
apparatus, defined by Eq.~(\ref{wiko}) for each possible value
of $x,$ and that $w(x,p;\th )$ of the particle given by
Eq.~(\ref{imam}). This factorization merely
expresses both the reduction of the initial state $w(x,p;0)$
into $w(x,p;\th )$ and the registration by the apparatus of the
classical random variable $x.$ 

The finite size of $d(\th )$ in Eq.~(\ref{werner1}) entails a
non-ideality of the measurement that we now consider. We shall
find it convenient to rewrite Eqs.~(\ref{pr1}), (\ref{werner1}) in two
alternative forms.
First we may use as indicated in Section 3.2b the density matrix representation only for the tested 
particle, while keeping the Wigner representation for the apparatus
(intermediate Wigner function):
\BEA
{\cal V}(X,P,x',x'';\th )=&&
\langle x'|\rho (\th ) | x'' \rangle
~\frac{\hbar}{\la}\exp \left[
-\frac{1}{2\la }\left (
X-\frac{g}{\alpha }~\frac{x'+x''}{2}
\right )^2-\frac{1}{2\la}P^2
-\frac{ g }{ \alpha }
\frac{{\rm i}}{\hbar} (x'-x'')P \right ],
\label{pr2}
\EEA
where the density matrix $\rho (\th )$ reduces to a nearly diagonal
form in $x$ as expressed by Eqs.~(\ref{shuravi}, \ref{khorezm}).
And finally the same expression can be presented in the complete 
density matrix representation:
\BEA
\label{werner200}
&&\langle X',x'|{\cal R}(\th )|X'',x''\rangle =\langle x'|\rho (\th )
| x'' \rangle ~\frac{1}{\sqrt{2\pi \la }}\exp \left[
-\frac{1}{2\la }\left ( \frac{X'+X''}{2}-\frac{g}{\alpha }~
\frac{x'+x''}{2} \right )^2 -\frac{\la }{2\hbar ^2}\left ( X'-X''-
\frac{g }{\alpha }(x'-x'') \right )^2 \right ].\nonumber\\
&&
\EEA
It is seen that the last small term $g(x'-x'')/\alpha$
in the r.h.s, of Eqs.~(\ref{pr2}) or (\ref{werner200}) quantifies
the entanglement, that is, the degree of
quantum correlations between the apparatus and the particle. Indeed,
if in Eq.~(\ref{werner200}) we neglect this factor, we again
find that the overall density matrix simply {\it factorizes} into the
contributions studied above separately for the apparatus and the
particle:
\BEA
\langle X',x'|{\cal R}(\th )|X'',x''\rangle \simeq
\langle X'|R_x|X'' \rangle~\langle x'|\rho (\th ) | x'' \rangle ,
\EEA
where we check, using $\lambda ~=~-\hbar T/\mu $ and
$\mu~=~-\hbar \alpha ,$ that
\BEA
\label{j}
R_x=
\frac{1}{\sqrt{2\pi \lambda}}\int \d X'~\d X''~
|X'\rangle\langle X''|~\exp \left[-\frac{1}{2\lambda}\left (
\frac{X'+X''}{2}-\frac{gx}{\alpha }
\right )^2-\frac{\lambda}{2\hbar ^2}\left (
X'-X''\right )^2\right ]
\EEA
is the gibbsian density operator of the apparatus for the
pointer variable $x~=~\left (x'+x''\right )/2$.
This factorization, together with the fact that $|\rho (\th ) |$ 
describes the reduced state of the tested particle, shows that the
program set up in the introduction by Eqs.~(\ref{1.1}~-~\ref{bars})
is achieved, and that our model describes an ideal measurement provided
the various conditions (\ref{kipchak2}~-~\ref{kipchak1}) are satisfied.

Let us turn to a more detailed discussion of the off-diagonal terms
in the overall density matrix (\ref{werner200}).
As we already discussed in section \ref{disco}, Eqs.~(\ref{pr2}, 
\ref{werner200}) are valid for the measurement of any hermitean 
operator $S$ of the tested particle, 
in particular those with a discrete spectrum. 
When dealing with this case, $\{ |x\rangle \}$ should be directly 
substituted by the eigenbase of $S$. Then the analogue of Eq.~(\ref{werner200})
reads:
\BEA
\label{shura}
\R (\th )= \sum _i p_i |s_i\rangle\langle s_i|\otimes R_i(\th )+
\sum _{i\not =k}
\exp \left [\frac{{\rm i}\xi (\th )}{2\hbar}(s_i^2-s_k^2)
-\frac{d(\th )}{2\hbar ^2}(s_i-s_k)^2 \right ]
\langle s_i|\rho (0)|s_k\rangle |s_i\rangle\langle s_k|
\otimes R_{ik},
\EEA
where $p_i=\langle s_i|\rho (0)|s_i\rangle$ is the initial distribution
of the measured quantity $S$ of the tested particle, and where
\BEA
&&R_{ik}=\frac{1}{\sqrt{2\pi \la }}\int \d X'~\d X''~
|X'\rangle\langle X''|~
\exp \left[
-\frac{1}{2\la }\left (
\frac{X'+X''}{2}-\frac{g}{\alpha }~\frac{s_i+s_k}{2}
\right )^2
-\frac{\la }{2\hbar ^2}\left (
X'-X''-\frac{g }{\alpha }(s_i-s_k)
\right )^2
\right ]
\nonumber\\ &&
\EEA
satisfies
\BEA
\label{snake}
{\rm tr}_{\rm A}R_{ik}=\exp \left [
-\frac{\la g^2}{\alpha ^2 }\frac{1}{2\hbar ^2}(s_i-s_k)^2 \right ],
\qquad {\rm tr}_{\rm A}R^2_{ik}=\frac{\hbar}{2\la }=
\frac{|\mu |}{2T}.
\EEA
For $i\not =k~R_{ik}$
has almost the same form as $ R_{ii}\equiv R_i$, 
given for the continuous case by Eq.~(\ref{j}), 
but with slightly shifted off-diagonal matrix elements.
Notice that this shift is due to entanglement between the apparatus and
the particle.

It is seen from Eq.~(\ref{shura}) that the off-diagonal terms of $\R$
are strongly suppressed with the exponential factor
\BEA\label{bogaz}
\exp \left [
-\frac{d(\th )}{2\hbar ^2}(s_i-s_k)^2 \right ].
\EEA
We again find the expected features for an ideal measurement, without
the difficulties of the continuous spectrum discussed in
Eq.~(\ref{foma}) and section \ref{tri}. 

In order to understand the above suppression in terms of observables,
let us imagine that one calculates the average in the state
(\ref{shura}) of some observable ${\cal F}$ with matrix elements $\langle
X'|{\cal F}_{ik}|X''\rangle $ in the Hilbert space of the
particle and the apparatus:
\BEA
\label{shura1}
\langle {\cal F}\rangle = \sum _i p_i 
~{\rm tr}_{\rm A}({\cal F}_{ii}R_i)+
\sum _{i\not =k}
\exp \left [\frac{{\rm i}\xi (\th )}{2\hbar}(s_i^2-s_k^2)
-\frac{d(\th )}{2\hbar ^2}(s_i-s_k)^2 \right ]
\langle s_i|\rho (0)|s_k\rangle 
~{\rm tr}_{\rm A}({\cal F}_{ki}R_{ik}).
\EEA
To ensure that the second sum in r.h.s. of Eq.~(\ref{shura}) is
non-negligible in the thermodynamic limit, one needs:
\BEA
\label{kolli}
|{\rm tr}_{\rm A}({\cal F}_{ki}R_{ik})|\propto
\exp \left [\frac{d(\th )}{2\hbar ^2}(s_i-s_k)^2 \right ]
\EEA
at least for one pair $(i\not =k)$. Using 
\BEA
|{\rm tr}_{\rm A}({\cal F}_{ki}R_{ik})|^2\le 
{\rm tr}_A{\cal F}_{ik}{\cal F}_{ki}~{\rm tr}_{\rm A} ~R^2_{ik}
\EEA
and Eq.~(\ref{snake}), one can write the condition (\ref{kolli}) as
\BEA
{\rm tr}_{\rm A}{\cal F}_{ik}{\cal F}_{ki}~\geq \frac{2T}{|\mu |}
\exp \left [\frac{d(\th )}{\hbar ^2}(s_i-s_k)^2 \right ].
\label{burundi}
\EEA
No general principle prohibits the existence of such an
observable ${\cal F}$ which will satisfy Eq.~(\ref{burundi}). However,
it is needless to mention that in the considered large $N$ limit it
would be quite pathological. So, under reasonable conditions
one only has the diagonal term in (\ref{shura1}).

The same conclusions as we just drew from the large size of $d(\th )$
do hold in the continuous case given by 
Eq.~(\ref{werner200}), except for divergences associated with the 
continuous spectrum, and we can rewrite Eq.~(\ref{werner200}) in the form
\BEA
\label{shura2}
\R (\th )= \int \d x~ p(x) |x\rangle\langle x|\otimes R_x(\th ),
\EEA where $p(x)=\langle x|\rho (0)|x\rangle$, which exhibits the form
required by the ideal measurement conditions.

Let us finally discuss for illustration two examples.

\subsubsection{Transformation of an eigenstate}

If the initial state $|x_1\rangle $ of the particle is an eigenstate of
coordinate, one has for the initial density matrix and Wigner function,
\BEA
\langle x'|\rho (0 ) | x'' \rangle
=\langle x'|x_1\rangle \langle x_1|x''\rangle ,
\qquad
w(x,p;0)=\delta (x-x_1).
\EEA
More precise normalization according to Eq.~(\ref{foma}) provides a small
width $\epsilon$ to the $\delta$-function in $w$ and multiplies it by
\BEA
\label{aharon}
\sqrt{8\pi\epsilon}~\exp\left ( -\frac{2\epsilon p^2}{\hbar ^2} \right).
\EEA

The state of the apparatus and particle for $\gamma \th \gg 1$ will be
\BEA
{\cal W}(X,P,x,p;\th ) = W_{x_1}(X,P)~\delta (x-x_1),
\EEA
where $W_{x_1}(X,P),$ given by Eq.~(\ref{wiko}), is the final Wigner function 
of the apparatus. As expected the measurement does not change the state of the
particle; it leaves the particle and the apparatus uncorrelated, apart from
the registration of the value $x_1$ in the latter.

\subsubsection{Decay of initial superpositions:
may Schr\"odinger kittens survive?}

Let the initial state of the particle be a superposition 
of two different eigenvectors $|x_1\rangle $ and $|x_2\rangle $ of the
coordinate operator, which appear with amplitudes $\varphi _1$,
$\varphi _2$. For simplicity we will take these amplitudes real. At the
initial time one has:
\BEA
\langle x'|\rho (0 ) | x'' \rangle =
\sum_{i,k=1}^2\varphi _i\varphi _k \langle x'|x_i\rangle \langle x_k|x''
\rangle .
\EEA
The corresponding Wigner function reads, within the regularization of 
Eq.~(\ref{foma}) which a provides a normalization factor (\ref{aharon}),
\BEA
w(x,p;0) =&& \sum_{i=1}^2 \varphi _i^2\delta (x-x_i)+
2\varphi _1\varphi  _2\delta \left (x-\frac{x_1+x_2}{2}
\right )\cos\frac{p(x_1-x_2)}{\hbar}.
\label{magadan}
\EEA
The physical interpretation of this formula is obvious.
The first two, incoherent terms refer to localized states at $x_1$ and
$x_2$. 
The cross term, which describes coherence, is localized half-way between
$x_1$ and $x_2$; through its oscillations it is associated with quantum
{\it interference}, a fact which is clear when one notices that it makes
the Wigner function alternatively positive and {\it negative} along the
line $y=(x_1+x_2)/2$. In other words, the initial state is highly
non-classical.

Using Eq.~(\ref{werner1}) one finds the common Wigner function
of the apparatus and the particle as
\BEA
\label{werner3}
{\cal W}(X,P,x,p;\th )=
\sum_{i=1}^2 \varphi _i^2 ~W_{x_i}(X,P) ~\delta (x-x_i)
+{\cal W}_{\rm if}(X,P,x,p;\th ),
\EEA
where the contribution from the interference term of $w(x,p;0)$
after integration over $p_0$ is:
\BEA
\label{werner20}
\W_{\rm if}(X,P,x,p;\th )=2\varphi _1\varphi _2
\exp\left [
-\frac{d(\th )}{2\hbar ^2}~(x_2-x_1)^2
\right ]
W_y(X,P)~
\delta \left (x-\overline{x}\right )
\cos \left[\frac{(x_1-x_2)}{\hbar}\left (
p+\frac{g}{\alpha }P-x\frac{g^2\th }{\alpha}
\right )\right ],
\EEA
with $\overline{x}=\left (x_1+x_2\right )/2$. 
The Wigner function (\ref{werner3})
is a sum of two contributions $W_{\rm m}$ and $W_{\rm if}.$
The first one, $W_{\rm m},$ is the expression (\ref{shura2})
describing an ideal measurement. It is a positive function and just
consists as expected of the incoherent mixture of two measured values
$x_1$ and $x_2$ with classical probabilities $\varphi _1^2$ and
$\varphi _2^2.$

The interference term $\W_{\rm if}$ is strongly suppressed due to the
factor $d(\th )$, which according to Eq. (4.52) yields in Eq.
(\ref{werner20}) an exponent of order $\gamma \th N^{3/2}$
provided $|x_2-x_1|$ is not very small. The disappearance of the
contribution of the interference term expresses that the initially
existing {\it Schr\"odinger cats} are automatically {\it suppressed},
so that a classical interpretation can be given to the final result of
the measurement.

For the continuous spectrum there can be cases where
$|x_2-x_1|\ll |x_1|$.
Since the initial superposition is then small, this situation can be
called {\it Schr\"odinger kitten}. As seen from Eq.~(\ref{werner20}),
the decay of such a state becomes less efficient as $|x_2-x_1|$
decreases. We may thus wonder whether Schr\"odinger kittens
could partly survive in a non-ideal measurement process. We
note, however, that the values $x_1$ and $x_2$ can be separated
in a measurement only if the transition probability of Eq.
(\ref{ichi}) is negligible, which requires
\BEA
\frac{\hbar g^2}{4T|\mu |}~\left (x_2-x_1\right )^2\gg 1.
\label{lili1}
\EEA
Since the exponent in the damping factor of Eq.~(\ref{werner20})
which characterizes the decoherence,

\BEA
\gamma \th
\frac{T^2}{\mu ^2}
\frac{\hbar g^2}{T|\mu |}\left (
x_2-x_1\right )
^2,
\label{lili2}
\EEA
is much larger by a factor of order $\gamma \th N$ than the exponent
(\ref{lili1}) which characterizes the robustness of the measurement,
even the weakest Schr\"odinger kittens disappear in any
measurement process.  Any kitten that can be detected by
distinguishing from each other the two interfering values of $x$,
has the same fate as a cat: it does not survive.

\renewcommand{\thesection}{\arabic{section}}
\section{Summary and conclusions}
\setcounter{equation}{0}

In the present paper we have studied a simple model in order to
get better insight on the question of quantum measurement. Our
purpose was to describe in full detail the dynamical process
which takes place during the measurement, due to the coupling
between the tested object and the apparatus. As usual in this
problem, we rely on the fact that the apparatus is a macroscopic
object, so as to ensure the rapid decoherence which is needed to
explain the classical nature of the interpretation of a
measurement. However, in our approach, this type of irreversible
behavior is merged with the idea that the apparatus should be
able to evolve indifferently towards different macroscopic
states. The selection of the outcome should be controlled by a
small interaction with the microscopic observed object. The
evolution of the apparatus should therefore be non ergodic,
which we realize by identifying the pointer variable with an
order parameter in a phase transition. The interaction with the
measured system behaves as a small source which drives the
actual value of the order parameter.

We wanted our model to fulfill the requirements on ideal
measurements listed by the end of the introduction. We also
wished to be able to produce, in a consistent framework and from
the first principles, using standard methods of quantum
statistical mechanics, a full solution for the equations of
motion which describe the dynamical process of measurement. This
led us to choose an extremely simple model. Our apparatus is a
non-interacting Bose gas in contact with a particle and energy
reservoir, which can undergo a Bose-Einstein condensation. The pointer
variable is the condensate density, which is sensitive to a
coupling of the lowest-energy level of the gas with the tested
microscopic system. To fix ideas we have chosen for this system a
one-dimensional particle, the position of which is to be
measured, but generalization to other systems is
straightforward. When the interaction between the system and the
apparatus is switched on, the condensate density relaxes to a
value in one-to-one correspondence with the possible values of
the position of the tested particle. We find that the randomness
of this position is directly reflected by the statistics of the
possible outcomes. The off-diagonal elements of the initial
density matrix of the particle are suppressed by the process,
and only classical probabilities enter the description of the correlations between the initial position of the particle and the pointer
variable.

The various parameters of the model can be tuned, so as to
explore the validity of the approximations which ensure that the
measurement is ideal. The following requirements, which were
expressed mathematically at the beginning of Section 4, are needed for
ideality.

a) The apparatus should be macroscopic. This large size plays a
double r\^ole. On the one hand it ensures though decoherence
the appearance of definite results in the measurement process. This
means that in its final state the overall system composed by the
apparatus and the tested system may be found in different mutually
exclusive states with probabilities given by the initial
distribution of the measured quantity. On the other hand, the
macroscopic number of condensed bosons ensures a robust and
accurate registration of the measured observable.

b) Before the interaction with the tested system starts to act, the
apparatus should be prepared in a state which is extremely sensitive
to this interaction with the tested microscopic system. When the
coupling is switched on at some initial time, the initial
state of the apparatus becomes thus unstable, and it relaxes to
another state determined by the measured object. This was achieved by
letting the initial number of condensed Bosons be already large (as
$\sqrt N$) but not yet macroscopic.

c) The relaxation time of the apparatus should be larger than
the dynamical time. 

d) The coupling constant $g$ should be finite, so that the
source term produces a macroscopic effect on the condensate
although the interaction term in the Hamiltonian is not
extensive.

e) The duration of the measurement should be larger than the
relaxation time of the apparatus.

f) The statistical distribution of the measured quantity, here
the position of the tested particle, should remain constant
during the whole measurement process.

The model explains the collapse of the state of the tested
particle as an effect of its coupling to the lowest level of the
Bose gas. The thermal bath acts only {\it indirectly}, through
the apparatus. We can therefore understand why this decoherence
process eliminates the off-diagonal elements of the density
matrix {\it in the $x$-basis} and not in another basis. Indeed,
the quantum noise due to the environment affects directly the
pointer variable $X$ of the apparatus, and the tested system
feels it only because it is coupled to $X$ through the term
$-g~x~X$ of the Hamiltonian. The association of the decoherence
with the measured observable is thus a natural outcome of the
model. Thus in our model the decoherence is determined by the interaction
between the tested system and the apparatus. This interaction is a
tunable property and can be controlled. Let us notice that in the
standard decoherence approach \cite{rev,omnes} this process 
depends on the interaction between the apparatus and its environment
which is a hardly controllable quantity by the very definition of (unobservable)
environment. 

Remarkably, the collapse of the state of the measured system takes place
at the very beginning of its interaction with the apparatus, over a time
$\tau = \hbar/\left ( TN^{1/4}\right )$ which has the usual features of
a decoherence time, proportional to $\hbar /T$ and small in the
thermodynamic limit of the apparatus. This time scale should be
contrasted to the much larger time scale $1/\gamma $ associated with
the relaxation of the apparatus. Once the state of the tested particle
is reduced, very little has yet been changed in the macroscopic
apparatus. It still takes a long time, of order $1/\gamma ,$ for the
apparatus to reach its new equilibrium position determined by the
system.

The reduction takes place for both the states of the tested
particle and the apparatus, which remain only coupled by
classical correlations at the end of the measurement. We have
estimated the order of magnitude of the corrections to this
ideal situation, and seen that they become extremely small as
$N$ increases. In particular the broadening of the density
matrix of the particle around its diagonal elements is of order
less than $N^{-3/2}.$ Related to this aspect is the suppression
of Schr\"odinger cats (interference effects of states located at
two different positions), and even of Schr\"odinger kittens
(similar states at two nearby positions), which cannot survive a
robust measurement.

If the tested system involves degrees of freedom other than the
one which is measured, our model shows that they remain
unaffected by the process. As expected for an ideal measurement,
the density matrix of the system is then simply projected
according to Eq. (\ref{1.5}) by the interaction with the
apparatus. In particular, in Einstein--Podolsky--Rosen setups,
our analysis confirms that the measurement of one particle
implies the collapse of the full system, thus also of constituents
that are spatially separated. This means that quantum information
is transferred at a large speed, which depends on the size of the
apparatus, and can thus be extremely large.

Although oversimplified, our model has many generic features
which are expected to occur in realistic measurement processes.
However, taking advantage of the lack of interactions in the Bose gas,
we have used an order parameter which is the density of the condensate.
This is a peculiar property, since in a real Bose gas only the
phase of the condensate, not its amplitude is an order
parameter. A drawback of this situation is the fact that the
back reaction of the apparatus on the system is stronger here
than expected for a truly ideal measurement: Indeed we have seen
that, apart from the reduction of the state to a nearly diagonal density
matrix, a large amount of momentum is transferred by
the apparatus to the particle. Another drawback of our model is
that it simulates, rather than completely involves non-ergodicity:
Although very sensitive to perturbations, our initial state is stable.
It would be desirable to work out a more elaborate model where the
initial state is metastable, and can be displaced towards several
possible truly equilibrium states characterized by an order parameter,
this displacement being controlled by the interaction with the system.

Other difficulties have been encountered above due to the
continuity of the spectrum of the measured quantity. For
instance, the large back reaction on the particle momentum is
related to this continuity. However, such difficulties are not an
artifact of our model as continuous spectra are known to cause
difficulties in many other circumstances.

\renewcommand{\thesection}{\arabic{section}}
\section*{Appendix A: Solution of equations of motion for an arbitrary
measured quantity}
\renewcommand{\theequation}{A.\arabic{equation}}
\setcounter{equation}{0}

Here we discuss the solution of equations of motion (\ref{vavilon1},
\ref{vavilon2}) for an interaction Hamiltonian
\BEA
\hat{H}_{\RS}=-g\HX\HS,
\EEA
where $\HS$ is an arbitrary hermitean operator, which can in particular have
discrete spectrum. In the present appendix we will distinguish operators
by a hat. Let us rewrite Eq.~(\ref{vavilon1}) in terms of $\HX$ and 
$\HP$:
\BEA
\label{vavilon8}
&& \frac{\d}{\d t}\HX = \alpha \HP -\gamma \HX + \hat{f}_X(t),\qquad
\hat{f}_X(t) = \sqrt{\hbar\gamma}(\hat{b}^{\dagger}(t)+\hat{b}(t)),
\\
\label{vavilon9}
&&\frac{\d}{\d t}\HP = -\alpha \HX -\gamma \HP +g\HS + \hat{f}_P(t),
\qquad
\hat{f}_P(t) = \ri\sqrt{\hbar\gamma}(\hat{b}^{\dagger}(t)-\hat{b}(t)).
\EEA
For a general operator of $\HS$ no Wigner function exists
in its standard sense, but for us it will be enough to operate with the
intermediate Wigner function $\V (X,P,s',s'')$ defined in section
\ref{rurk}. Recall that this object is a Wigner function with respect to
the lowest level of the apparatus (here the corresponding
Wigner function can be defined), but is a density matrix in the 
eigenrepresentation $\{ |s\rangle \}$ of $\HS $.

To write the equations of motion in terms of $\V$ we will use a method
described in \cite{ABN}, which replaces the operator equations (A.2),
(A.3) by the stochastic equation:
\BEA
\dt \tilde{\V}(X,P,s',s'')
=&&-\dX ([\alpha P-\gamma X]\tilde{\V})+\dP ([\alpha X+\gamma P]\tilde{\V})
\nonumber\\
&&-\dX (f_X\tilde{\V})-\dP (f_P\tilde{\V})
\nonumber\\
&&-\frac{s'+s''}{2}g\dP \tilde{\V}
+\frac{\ri g}{\hbar}X(s'-s'')\tilde{\V}.
\label{blot}
\EEA
Here $f_X(t)$ and $f_P(t)$ behave as classical noises, which have
exactly the
same average and autocorrelation as the corresponding quantum quantities
$\hat{f}_X(t)$ and $\hat{f}_P(t)$ after the symmetrization of
Eq.~(2.42). The true intermediate Wigner function $\V$ is
obtained from $\tilde{\V}$ by averaging with respect to these
noises $f_X(t)$ and $f_P(t)$,
\BEA
\V (X,P,s',s'')=\langle\tilde{\V}(X,P,s',s'')\rangle
\EEA
The first line in the r.h.s. of Eq.~(\ref{blot}) is the standard drift 
contribution of the Liouville-Wigner equation. The last line in this equation
is as well explained rather simply: This is just the result of the Wigner
transformation (\ref{rhoW}) which was applied for the apparatus to the
corresponding term $(\ri/\hbar )\langle s'|[H,\rho ]|s''\rangle$
in the density matrix representation.

Now it is easy to see by direct substitution that the solution of
Eq.~(\ref{blot}) reads:
\BEA
\tilde{\V}(X,P,s',s'')=
\exp \left[
\frac{\ri g}{\hbar}(s'-s'')\int _0^t\d t' X(t')
\right ]\delta (X-X(t))\delta (P-P(t)),
\EEA
where $X(t)$ and $P(t)$ are the solution of the following c-number
equations

\BEA
\label{vavilon10}
&& \frac{\d}{\d t}X = \alpha P -\gamma X + f_X(t),
\\
\label{vavilon11}
&&\frac{\d}{\d t}P = -\alpha X -\gamma P + g\frac{s'+s''}{2}+ f_P(t).
\EEA
This means that one has the following solution for $\V$:
\BEA
\V(X,P,s',s'')=\int \d X_0\d P_0\left \langle
\exp \left[
\frac{\ri g}{\hbar}(s'-s'')\int _0^t\d t' X(t')
\right ]\delta (X-X(t))\delta (P-P(t))\right\rangle 
W(X_0,P_0)\langle s'|\rho (0)|s''\rangle,
\label{gumo}
\EEA
where $W(X_0,P_0)$ is the initial Wigner function of the lowest level,
and $X_0=X(0)$, $P_0=P(0)$ are initial values of $X(t)$, $P(t)$
which are inherent in Eqs.~(\ref{vavilon10}, \ref{vavilon11}).
Since the exact solution of these equations is available in the form
of Eqs.~(\ref{tulukan1}, \ref{tulukan2}), a little patience is 
sufficient to verify that $\V(X,P,s',s'')$ in Eq.~(\ref{gumo}) 
coincides with the intermediate Wigner function corresponding to 
Eqs.~(\ref{mr}, \ref{kr}) provided that one makes the identification 
$s'\to x'$, $s''\to x''$.

\renewcommand{\thesection}{\arabic{section}}
\section*{Appendix B: The state of the particle after measurement}
\renewcommand{\theequation}{B.\arabic{equation}}
\setcounter{equation}{0}

Here we discuss the post-measurement evolution of the state of the
particle. We will assume that at time $t=\th $ the interaction between
the particle and the apparatus has been instantaneously switched off. 
The switching is needed to ensure that the measurement will remain
ideal, whereas its instantaneous character is taken for simplicity.
Indeed, if the tested particle is interacting with the apparatus long
enough, its coordinate will start to change due to its own Hamiltonian
$H_{\rm S}$,
as well as due to interaction with the apparatus. Since the apparatus is
itself interacting with the bath, sufficiently long interaction of the
particle and apparatus will finally lead to relaxation of
the particle towards certain steady state, which is independent of 
its initial state. This will violate the condition of ideality.

Therefore, the time $\th$ was assumed to be much larger than $1/\gamma$, 
so that the apparatus has enough time to relax to its stationary state and 
monitor the results of measurement. On the other hand, $\th $ was assumed
to be small enough so that effects connected with change of 
$x$ are not yet relevant. For $t\ge \th $ the tested particle thus
follows its free evolution, which is described by free Heisenberg
equations:
\BEA
&& p(t)=p(\th ),
\\
&& x(t)=x(\th )+\frac{t-\th }{m}p(\th )
\EEA
Due to the instantaneous character of the switching, $p(\th )$, $x(\th )$
are those operator values which the particle reached during the
interaction with the apparatus.

The dynamics from $t=0$ to $t=\th $ is described by Eq.~(\ref{molla}),
and for $t>\th $ one has
\BEA
w(x,p;t) = \int \d \Bx ~\d \Bp ~w(\Bx,\Bp ,\th )
\delta(p-\Bp )~\delta \left (x-\Bx -\frac{t-\th }{m}\Bp\right ).
\label{re-molla}
\EEA
Let us now investigate $w(x,p;t)$ at $t>\th $ for the initial state at $t=0$
given by Eq.~(\ref{magadan}):

\BEA
w(x,p;t) = &&
\sum _{k=1}^2 \varphi _i^2\delta \left (x-x_i-\frac{t-\th }{m}p \right )
\nonumber\\
+&&
2\varphi _1\varphi _2\delta \left (x-\frac{t-\th }{m}p-\frac{x_1+
x_2}{2}\right )
\exp \left [-\frac{d(\th )(x_1-x_2)^2}{2\hbar ^2}\right ]
\cos \left [\frac{(x_1^2-x_2^2)}{2\hbar }(p-\xi (\th )x) \right ]
\label{kolima}
\EEA
It is seen that the last contribution to this equation is due to
incomplete reduction. If it is not suppressed totally in the course of
the measurement (because $N$ is not sufficiently large), it will persist
during further evolution, and might in principle be observed.

\renewcommand{\thesection}{\arabic{section}}
\section*{Appendix C: Motion of the particle during measurement}
\renewcommand{\theequation}{C.\arabic{equation}}
\setcounter{equation}{0}

The purpose of the present Appendix is to discuss what happens
if the change with time of the measured quantity $x$ is not negligible,
and if it cannot be treated as a constant of motion. In practice the
conservation of the measured quantity is ensured only over short times.
However the duration $\th $ of the interaction with the apparatus should
be sufficient so as to ensure registration, and there will arise a
source of non-ideality.
Our purpose here is not to develop a full account of this
non-ideality, but just to display on which time-scales its presence is
not relevant. 

We will investigate Eqs.~(\ref{vavilon1}, \ref{vavilon6}, 
\ref{vavilon7}) on times where changes of the measured quantity $x$ 
become noticeable. 
To keep the situation free of instabilities, we will make a natural 
assumption that the tested particle is subjected to a confining
potential. For simplicity this potential will be taken to be harmonic: 
$V(x)=m\omega _0^2x^2/2$.

\subsection*{Dynamics}
The general solution of Eqs.~(\ref{vavilon6}, \ref{vavilon7})
is obtained with help of Laplace transformation. Recall
the following standard relations between functions $A(t)$, $B(t)$ and 
their Laplace-transforms 
$\hat A(s)=\int_0^\infty\d t\,e^{-st}A(t)$ and $\hat B(s)$ denoted
in this appendix with a hat:
\BEA
{\cal L}\left \{
\int _0^t\d t' A(t-t')B(t')
\right \}=\hat{A}(s)\hat{B}(s),
\qquad
{\cal L}\left \{\dot{A}\right \}=-A(0)+s\hat{A}(s),
\EEA
where $\dot{A}=\frac{\d}{\d t}A$.
Thus, the solution of Eqs.~(\ref{vavilon6}, \ref{vavilon7}) reads
\BEA
&&
\hat{x}(s)=\frac{1}{m}\hat{f}(s)\hat{\eta}(s)+(\dot{x}(0)+sx(0))
\hat{f}(s),
\\
&&
\hat{f}(s)=\frac{1}{s^2+\omega _0^2-\frac{1}{m}\hat{\chi}(s)},
\EEA
where
\BEA
\hat{\chi}(s)  = g^2~\frac{\alpha }{\alpha ^2+(\gamma +s)^2}
\EEA
is the Laplace transform of $\chi (t)$ (see Eq.~(\ref{hh=})). Finally
one has
\BEA
\label{khan1}
&&x(t)=x(0)\dot{f}(t)+\frac{1}{m}p(0)f(t)+\frac{1}{m}\int_0^{t}\d t'
f(t-t')\eta (t'),
\\
&&p(t)=p(0)\dot{f}(t)+mx(0)\ddot{f}(t)+\int_0^{t}\d t'
\dot{f}(t-t')\eta (t'),
\label{khan2}
\EEA
where $f(t)$ is the inverse Laplace transform of $\hat{f}(s)$.

Substituting Eq.~(\ref{khan1}) into Eq.~(\ref{plumbum}), we
obtain the solution of the Heisenberg equation for the pointer variable:
\BEA
X(t)=gx(0) \int ^t_0dt'{\rm e}^{-\gamma t'} \sin \alpha t' \dot{f}
(t-t')+\frac{gp(0)}{m}\int ^t_0dt'{\rm e}^{-\gamma t'}\sin
\alpha t'f(t-t')
+\frac{1}{g}{\cal{L}}^{-1}\left \{ 
\left (\frac{\hat{\chi}}{\it m}\hat{f} +1
\right )
{\hat{\eta}}\right \} .
\label{balanda}
\EEA
Its last term, where $\eta $ is given by Eqs.~(\ref{etat=}--2.36)
describes the effect of the noise $b,b^{\dagger }$ and the
remanence of the initial conditions $X(0),P(0)$ of the
apparatus. We recover Eq.(\ref{tulukan1}) in the large $m$ limit,
for which $f(t)=t$ if $\omega_0t\ll 1.$ However, instead of
being controlled by $x(0)$ only, the order parameter $X(t)$ now
depends on the initial state of the particle through both $x(0)$
and $p(0).$ Since due to the uncertainty relation
$\langle p(0)^2\rangle\langle x(0)^2\rangle \ge \hbar ^2/4$ these
quantities cannot have definite values simultaneously,
$\langle X(t)\rangle$ is always fluctuating with the initial state of
the particle. This violates conditions (\ref{1.2}, \ref{1.3}) according
to which if the tested system starts its evolution from one of the
eigenstates of the measured quantity, then the result displayed by the
apparatus can be definite. Moreover Eq. (\ref{balanda}) shows that
the statistics of $X(t)$ is governed by the statistics of
$x(0),p(0),$ or equivalently by the full density operator $\rho (0)$
of the particle at the beginning of the measurement,
including off-diagonal elements. It is the disappearance of the
term in $p(0)$ in Eq.(\ref{tulukan1}) which allows $X(0)$ at the
end of the measurement to depend only on the diagonal elements
$\langle x|\rho (0)|x\rangle .$ As we shall see
below, this occurs for $m\gg g^2\th ^3,$ that is, for a
sufficiently short duration of the measurement.
Once more this shows that the short-time limit is a necessary condition
for (nearly) ideal measurement of the coordinate.

Remember that the Heisenberg equation (\ref{vavilon1}) for $X(t)$
depended only on the position operator $x(t),$ not on the
momentum operator $p(t).$ Its solution given by Eq.~(\ref{plumbum})
expresses the action of the particle on the apparatus as a
{\it memory effect} depending on $x(t)$ at earlier times. Using the
equation of motion (\ref{khan1}) for $x(t),$ we have here re-expressed
$X(t)$ in terms of $x(0)$ and $p(0).$ The momentum of the particle
thus came out from the elimination of the history of $x(t).$

\subsection*{Short-time expansion}

To investigate the short-time behavior of the tested particle in more
detail, we adopt the following large-mass approximation:
\BEA
&&\hat{f}(s) = \frac{1}{s^2+\omega _0^2}
+\frac{1}{(s^2+\omega _0^2)^2}\frac{\hat{\chi}(s)}{m},\\
&&f(t)=\frac{ \sin \omega _0t}{\omega _0}
+\frac{\Omega ^3}{2\omega _0^3}\int _0^t\d t'
[\sin (\omega _0t')-t'\omega _0\cos \omega _0t']\sin (\alpha [t-t'])
e^{-\gamma (t-t')},
\label{rashid}
\EEA
where
\BEA
\label{sasanid}
\Omega  =\left ( 
\frac{g^2}{m}
\right )^{\frac{1}{3}}
\EEA
is the characteristic frequency connected with the mass of the particle 
and the interaction with the apparatus. This frequency will be assumed
to be the smallest characteristic frequency in our problem.
Eq.~(\ref{rashid}) is a short-time expansion, which is valid for
\BEA
t\ll \frac{1}{\Omega }=\left (\frac{m}{g^2}\right )^{\frac{1}{3}}.
\label{baclan}
\EEA
Expansion (\ref{rashid}), when substituted into Eq.~(\ref{khan1}),
produces the coordinate as a sum of two terms: The first one is due to
the free motion in the potential $V(x)$, and the second one represents
a deterministic correction arising from interaction with the apparatus.
The interaction with the apparatus should be switched off before this
term becomes comparable with the first term in Eq.~(\ref{rashid}).
In particular, if $t$ is so small as $t\ll 1/\omega _0$ we obtain:
\BEA
\label{bek}
f(t)=t
+\frac{\Omega ^3}{6}\int _0^t\d t' t'^3
\sin (\alpha [t-t'])
{\rm e}^{-\gamma (t-t')}.
\label{dustum}
\EEA
When this equation is substituted into Eqs.~(\ref{khan1}, \ref{khan2}),
it is seen that in all terms besides the second term in the r.h.s. of
Eq.~(\ref{khan2}) the correction to $f(t)=t$ in Eq.~(\ref{bek}) can be
dropped under the condition (\ref{baclan}). If one will take
additionally: 
$x(0)\gg p(0)t/m$, Eqs.~(\ref{khan1}, \ref{khan2}, \ref{bek})
produce Eq.~(\ref{hu}). 

It is important to notice that in this approximation there is a
deterministic influence of the bath on the momentum. This is just
friction, which may enhance or reduce $\langle p\rangle .$ However,
there is no such a systematic influence on the coordinate. It was
therefore legitimate to assume in the bulk of this paper that
in the Heisenberg representation we have $x(t)\simeq x(0)$ in the time
interval $0,\th $ in spite of the large value reached by
$\langle p\rangle$ at the time $\th $ due to friction (Section
4.2b).

\subsection*{Relaxation of the tested particle}

Although Eq.~(\ref{rashid}) has been sufficient for our purposes, we
will mention here how the full $f(t)$ behaves. This will allow us
to understand the long-time behavior of the tested particle.
One has to obtain the roots $\gamma _1,\gamma _2,\gamma _3,\gamma _4$ 
of the following equation
\BEA
[\alpha ^2+(\gamma _k -\gamma )^2][\gamma ^2_k+\omega _0^2]-\alpha
\Omega ^3 = 0
\EEA
which provide
\BEA
f(t)=\sum _{k=1}^4 A_k e^{-\gamma _k t},\qquad 
A_k ={\rm Lim}_{s\to -\omega_k}(s+\gamma _k)\hat{f}(s).
\EEA
Using the smallness of $\Omega $ the above roots can be obtained
approximately as
\BEA
\label{batono1}
&& \gamma _1= \gamma +i\alpha +\delta _1,\qquad
\gamma _2=\gamma _1^*,\\
&& \gamma _3=i\omega _0+\delta _2,\qquad
\gamma _4=\gamma _3^*,
\label{batono2}
\EEA
where, after using $\alpha \gg \gamma,~\alpha \gg \omega _0,$ we
find:
\BEA
&&\delta _1=\frac{\Omega ^3}{2{\rm i}}~
\frac{1}{(\gamma +{\rm i} \alpha )^2+\omega _0^2}+{\cal O}(\Omega ^6)
\simeq
-\frac{\Omega ^3 \gamma }{\alpha ^3}+
\frac{{\rm i}\Omega ^3}{2\alpha ^2},\\
&&\delta _2=
\frac{\alpha \Omega ^3}{2{\rm i}\omega _0}~
\frac{1}{({\rm i}\omega _0-\gamma )^2
+\alpha ^2}+{\cal O}(\Omega ^6)\simeq \frac{\Omega ^3\gamma
}{\alpha ^3}-\frac{{\rm i}\Omega ^3}{2\alpha \omega _0}.
\EEA
As seen from Eqs.~(\ref{batono1}, \ref{batono2}) the particle has
two relaxation times:
$1/\gamma$ (which is also the relaxation time of the apparatus) and
a much longer one $1/({\rm Re}~\delta _2)=\alpha ^3/\Omega ^3\gamma .$
They are widely separated since $\alpha ^3/\Omega ^3\gg 1$. After the
time $\alpha ^3/\Omega ^3\gamma =\alpha ^3g^2/m\gamma $ all information
about the initial state will be forgotten by the particle if it still
interacts with the apparatus, and it will relax to its equilibrium state
imposed by the bath temperature.

\renewcommand{\thesection}{\arabic{section}}
\section*{Appendix D: Weak-coupling Langevin equation}
\renewcommand{\theequation}{D.\arabic{equation}}
\setcounter{equation}{0}

In the present appendix we will derive from a consistent Hamiltonian
formulation the Langevin equation (2.21) for the apparatus-bath dynamics.
More information on weakly coupled (weakly damped) dissipative systems can
be found in [24].

For simplicity we deal only with the lowest energy level of the apparatus,
the Hamiltonian of which reduces to the chemical potential term $-\mu
\,a^{\dagger }a=\hbar \alpha \,a^{\dagger }a;$ as in Eq. (2.28) we drop the
index $0$. The dynamics of the excited state operators $a_{i}$ would be
obtained by substituting $\alpha +\omega _{i}$ to $\alpha $. The considered
mode is coupled to the thermal bath, which is also a non-interacting Bose
gas having single-particle levels $m$ with creation operators $\xi
_{m}^{\dagger }$ and energies $\hbar \Omega _{m}$ (including the chemical
potential of the bath). This bath is equivalent to the dense set of harmonic
oscillators usually considered in quantum Brownian motion [20, 24, 25], but
here the coupling should account for transfers of bosons between the bath
and the apparatus proper, with conservation of the total particle number. We
can thus describe the apparatus with its bath by the total Hamiltonian
\begin{equation}
H_{\rm AB }=\hbar \alpha \,a^{\dagger }a+\sum_{m}\hbar \Omega _{m}\,\xi
_{m}^{\dagger }\,\xi _{m}+\sum_{m}\,\hbar \left( c_{m}a^{\dagger }\xi
_{m}+c_{m}^{\ast }\xi _{m}^{\dagger }a\right) .  \label{D.1}
\end{equation}

The resulting equations of motion in the Heisenberg picture are:
\begin{equation}
\dot{a}=-\ri\alpha\, a -\ri\sum_{m}c_{m}\xi _{m},  \label{D.3}
\end{equation}
\begin{equation}
\dot{\xi}_{m}=-\ri\Omega _{m}\xi _{m}-\ri c_{m}^{\ast }a. 
\label{D.4a}
\end{equation}

\noindent%
The initial state of the apparatus and the bath is assumed to be factorized
at some remote initial time $t_{0}.$ At that time the bath was in
equilibrium at temperature $\beta ^{-1}$ while the apparatus was in an
arbitrary state. Explicit integration of Eq. (D.3), through%
\begin{equation}
\xi _{m}\left( t\right) =\xi _{m}\left( t_{0}\right) \re^{-\ri
\Omega _{m}\left( t-t_{0}\right) }-\ri c_{m}^{\ast }\int_{t_{0}}^{t}
\d t^{\prime }\,a\left( t^{\prime }\right) \re ^{\ri 
\Omega _{m}\left( t-t^{\prime }\right) },  \label{D.4}
\end{equation}%
allows us to eliminate the bath and to write a closed, exact equation of
motion for the operator $a.$ In the weak-coupling regime it is convenient to
go the rotating frame by means of the transformation%
\begin{equation}
\tilde{a}\left( t\right) =a\left( t\right) \re ^{\ri \alpha
\left( t-t_{0}\right) }.  \label{D.5}
\end{equation}

\noindent%
From Eqs. (D.2), (D.4) and (D.5) we obtain%
\begin{equation}
\frac{\d \tilde{a}\left( t\right) }{\d t}=-\sum_{m}\left|
c_{m}\right| ^{2}\int_{t_{0}}^{t}\d t^{\prime }\,\tilde{a}\left(
t^{\prime }\right) \re ^{\ri \left( \alpha -\Omega _{m}\right)
\left( t-t^{\prime }\right) }+\zeta \left( t\right) ,  \label{D.6}
\end{equation}%
where $\zeta \left( t\right) $ appears as a gaussian quantum noise defined
by 
\begin{equation}
\zeta \left( t\right) =-\ri \sum_{m}c_{m}\xi _{m}\left( t_{0}\right) 
\re ^{\ri \left( \alpha -\Omega _{m}\right) \left(
t-t_{0}\right) }.  \label{D.7}
\end{equation}%
This noise is characterized by the properties%
\begin{equation}
\lbrack \zeta \left( t\right) ,\zeta ^{\dagger }\!\left( t^{\prime }\right)
]=\sum_{m}\left| c_{m}\right| ^{2}\re ^{\ri \left( \alpha
-\Omega _{m}\right) \left( t-t^{\prime }\right) },\quad \quad \left[ \zeta
\left( t\right) ,\zeta \left( t^{\prime }\right) \right] =0,  \label{D.8}
\end{equation}%
\begin{equation}
\left\langle \zeta ^{\dagger }\!\left( t^{\prime }\right) \zeta \left(
t\right) \right\rangle =\sum_{m}\frac{\left| c_{m}\right| ^{2}\re ^{%
\ri \left( \alpha -\Omega _{m}\right) \left( t-t^{\prime }\right) }}{%
\re ^{\beta \hbar \Omega _{m}}-1},\quad \quad \left\langle \zeta
\left( t\right) \zeta \left( t^{\prime }\right) \right\rangle =0. \label{D.9}
\end{equation}

We are interested in the thermodynamic limit for the bath, where the number
of bath modes goes to infinity while their frequencies $\Omega _{m}$ tend to
a continuum. For simplicity, we assume that the values $\Omega _{m}$ have a
constant spacing $\Delta \rightarrow 0,$ and that the coupling $\left|
c_{m}\right| $ is a vanishingly small constant:%
\begin{equation}
\left| c_{m}\right| =\sqrt{\frac{\gamma \Delta }{\pi }}.  \label{D.10}
\end{equation}%
We then have%
\begin{equation}
\sum_{m}\left| c_{m}\right| ^{2}\re ^{\ri \left( \alpha -\Omega
_{m}\right) \left( t-t^{\prime }\right) }=2\gamma \delta \left( t-t^{\prime
}\right) .  \label{D.11}
\end{equation}%
(If the spectrum of $\Omega _{m}$ is bounded, the $\delta $-function stands
for a narrow peak, with a width smaller than all the characteristic times of
the problem if the range of $\Omega _{m}$ is sufficient.) Eq. (D.11) ensures
that the retardation effect can be neglected in Eq. (D.6), which becomes:%
\begin{equation}
\frac{\d \tilde{a}\left( t\right) }{\d t}=-\gamma \tilde{a}%
\left( t\right) +\zeta \left( t\right) ,  \label{D.12}
\end{equation}%
and that the properties of the noise are simplified into%
\begin{equation}
\lbrack \zeta \left( t\right) ,\zeta ^{\dagger }\left( t^{\prime }\right)
]=2\gamma \delta \left( t-t^{\prime }\right) ,\quad \quad \left[ \zeta
\left( t\right) ,\zeta \left( t^{\prime }\right) \right] =0  \label{D.13}
\end{equation}%
\begin{equation}
\left\langle \zeta ^{\dagger }\left( t^{\prime }\right) \zeta \left(
t\right) \right\rangle =\frac{\gamma }{\pi }\int \frac{\d \omega \,%
\re ^{\ri \omega \left( t-t^{\prime }\right) }}{\re %
^{\beta \hbar \left( \alpha -\omega \right) }-1},\quad \quad \left\langle
\zeta \left( t\right) \zeta \left( t^{\prime }\right) \right\rangle =0. 
\label{D.14}
\end{equation}

Apart from the very short time-scale involved in the noise $\zeta \left(
t\right) ,$ two time-scales, $\alpha ^{-1}$ and $\gamma ^{-1},$ enter the
dynamical equations (D.5), (D.12) for the operator ${a(t)}$ generated
by the apparatus-bath Hamiltonian $H_{\mathrm{A,B}}$. Other, longer
time-scales will also be induced by the interaction with the tested system.
We assume that the \textit{dynamical frequency} $\alpha $ \textit{is the
largest }characteristic frequency of the problem. The transformation (D.5)
then acounts for the fast, oscillatory motion of ${a(t)}$. The
evolution of $\tilde{a}\left( t\right) $ takes place on larger time-scales,
of order $\gamma ^{-1}$ or more. We can thus expand Eq. (D.14) for large $%
\alpha $ according to 
\begin{eqnarray}
\left\langle \zeta ^{\dagger }\left( t^{\prime }\right) \zeta \left(
t\right) \right\rangle  &\approx &\frac{\gamma }{\pi }\int \d \omega
\,\re ^{\ri \omega \left( t-t^{\prime }\right)
}\sum_{n=0}^{\infty }\frac{\left( -\omega \right) ^{n}}{n!}\left( \frac{%
\mathrm{d\,\,\,\,}}{\d \alpha }\right) ^{n}\frac{1}{\re ^{\beta
\hbar \alpha }-1}  \nn 
\\
&=&2\gamma \sum_{n=0}^{\infty }\frac{\ri ^{n}}{n!}\delta ^{\left(
n\right) }\left( t-t^{\prime }\right) \left( \frac{\mathrm{d\,\,\,\,}}{%
\d \alpha }\right) ^{n}\frac{1}{\re ^{\beta \hbar \alpha }-1}. 
\end{eqnarray}%
Since this expression will be integrated over a function of time which
varies on the scale $\gamma ,$ its successive terms yield an expansion in
powers of $\gamma /\alpha ,$ and we can thus replace Eq.(D.14) in the limit $%
\alpha \gg \gamma $ by 
\begin{equation}
\left\langle \zeta ^{\dagger }\left( t^{\prime }\right) \zeta \left(
t\right) \right\rangle =2\gamma \frac{\delta \left( t-t^{\prime }\right) }{%
\re ^{\beta \hbar \alpha }-1},\quad \quad \left\langle \zeta \left(
t\right) \zeta \left( t^{\prime }\right) \right\rangle =0.  \label{D.16}
\end{equation}

If the noise is redefined as 
\begin{equation}
b_{0}\left( t\right) =\frac{1}{\sqrt{2\gamma }}\zeta \left( t\right) \mathrm{%
e}^{-\ri \alpha \left( t-t_{0}\right) },  \label{D.17}
\end{equation}%
Eq. (D.12) reduces to the equation of motion (2.21) for $a_{0}\left(
t\right) ,$ and Eqs. (D.13), (D.16) coincide with the conditions (2.23),
(2.24), (2.25) on $b_{0}\left( t\right) .$ We can likewise recover Eqs.
(2.21--2.25) for $\mathit{i}$\textrm{\ }and $k\neq 0$ by adding $\omega _{k}$
to $\alpha $ and by coupling all the modes $\mathit{i}$ of the apparatus
with the modes $m$ of the bath.

If we had not used the approximation $\alpha \gg \gamma ,$ we would have
obtained the evolution of the average particle number $n_{0}\left( t\right)
\equiv \left\langle a^{\dagger }\left( t\right) a\left( t\right)
\right\rangle $ in the lowest level of the apparatus by integration of Eqs.
(\ref{D.5}), (\ref{D.12}), using (\ref{D.14}) instead of (\ref{D.16}). 
This yields%
\begin{equation}
n_{0}\left( t\right) =n_{0}\left( t_{0}\right) \re ^{-2\gamma \left(
t-t_{0}\right) }+\frac{\gamma }{\pi }\int \frac{\d \omega }{\left(
\omega -\alpha \right) ^{2}+\gamma ^{2}}\left| 1-\re ^{-\left( \gamma +%
\ri \alpha -\ri \omega \right) \left( t-t_{0}\right) }\right|
^{2}\frac{1}{\re ^{\beta \hbar \omega }-1},  \label{D.18}
\end{equation}%
instead of Eq. (2.27). In particular, we find for arbitrary $\gamma /\alpha $
that $n_{0}\left( t\right) $ relaxes for large times towards%
\begin{equation}
n_{0}\left( t\right) \rightarrow \frac{\gamma }{\pi }\int \frac{\d %
\omega }{\left( \omega -\alpha \right) ^{2}+\gamma ^{2}}\,\frac{1}{\re %
^{\beta \hbar \omega }-1}. \label{D.19}
\end{equation}%
It is only for $\gamma \ll \alpha $ that Eq. (\ref{D.18}) 
reduces to Eq. (2.27),
and that $n_{0}\left( t\right) $ reaches the expected equilibrium Bose
factor $1/\left( \re ^{\beta \hbar \alpha }-1\right) $ for large times.

Altogether the weak-coupling condition $\gamma \ll \alpha ,$ which will be
enforced throughout this paper, ensures that the quantum Langevin equation
(2.21) can be obtained from the apparatus-bath Hamiltonian (D.1). Owing to
this condition, while the coupling parameter $\gamma $ governs the \textit{%
dynamics} of the apparatus during its relaxation to equilibrium, the \textit{%
equilibrium} properties of the apparatus remain unaffected by the presence
of the bath. Note finally that the simplifying assumptions we made on the
bath (equally spaced levels $\Omega _{m},$ constant $\left| c_{m}\right| )$
become irrelevant in the weak-coupling limit.

\acknowledgments R. B. is grateful for discussion with M. Gaudin.
Th.M. N. acknowledges hospitality during a visit to the CEA Saclay.
A.E. A. acknowledges support by NATO.

\end{document}